\documentclass[10 point, journal, comsoc]{IEEEtran}
\usepackage{amsmath,amsfonts}
\usepackage{algorithm}  
\usepackage{algorithmic}  
\usepackage{array}
\usepackage{color}
\usepackage{color}
\usepackage[cmyk]{xcolor}
\usepackage{booktabs}
\usepackage[caption=false,font=footnotesize,labelfont=rm,textfont=rm]{subfig}
\usepackage{textcomp}
\usepackage{stfloats}
\usepackage{url}
\usepackage{verbatim}
\usepackage{graphicx}
\usepackage{cite}
\usepackage{tabularx}
\usepackage{caption}
\usepackage{threeparttable}
\usepackage{amsthm,amsmath,amssymb}
\usepackage{mathrsfs}
\usepackage[switch]{lineno}
\usepackage{url}
\usepackage{fancyhdr}
\begin{document}

\title{Task Containerization and Container Placement Optimization for MEC: A Joint Communication and Computing Perspective}

\author{Ao~Liu, Shaoshi~Yang, Jingsheng~Tan, Zongze~Liang, Jiasen~Sun, Tao~Wen, and Hongyan~Yan
\thanks{This work was supported in part by the Beijing Municipal Natural Science Foundation under Grant L202012, and in part by the BUPT-CMCC Joint Research Center under Grant A2022122. \textit{(Corresponding author: S. Yang)}.  %in part by the Fundamental Research Funds for the Central Universities under Grant 2020RC05, in part by the Engineering and Physical Sciences Research Council project EP/P003990/1 (COALESCE), and in part by the European Research Council's Advanced Fellow Grant QuantCom (Grant No. 789028)

A. Liu, S. Yang, J. Tan, Z. Liang and J. Sun are with the School of Information and Communication Engineering, Beijing University of Posts and Telecommunications, and also with the Key Laboratory of Universal Wireless Communications, Ministry of Education, Beijing 100876, China (E-mail: \{ao.liu, shaoshi.yang\}@bupt.edu.cn).

T. Wen and H. Yan are with the China Mobile
Research Institute, Beijing 100053, China. 

Published on \textit{Processes} \textbf{2023}, \textit{11(5)}, article number: 1560. https://doi.org/10.3390/pr11051560}
}% 

%\markboth{Published in Processes 2023, 11, 1560. doi: 10.3390/pr11051560}%
%{Shell \MakeLowercase{\textit{et al.}}: Bare Demo of IEEEtran.cls for IEEE Communications Society Journals}

\maketitle

% As a general rule, do not put math, special symbols, or citations
\begin{abstract}
Containers are used by an increasing number of Internet service providers to deploy their applications in multi-access edge computing (MEC) systems. Although container-based virtualization technologies significantly increase application availability, they may suffer expensive communication overhead and resource use imbalances. However, so far there has been a scarcity of studies to conquer these difficulties. In this paper, we design a workflow-based mathematical model for applications built upon interdependent multitasking composition, formulate a multi-objective combinatorial optimization problem composed of two subproblems---graph partitioning and multi-choice vector bin packing, and propose several joint task-containerization-and -container-placement methods to reduce communication overhead and balance multi-type computing resource utilization. The performance superiority of the proposed algorithms is demonstrated by comparison with the state-of-the-art task and container scheduling schemes.
\end{abstract}

% Note that keywords are not normally used for peer review papers.

% \fi
%
% For peer review papers, this IEEEtran command inserts a page break and
\IEEEpeerreviewmaketitle

\section{Introduction}

\IEEEPARstart{T}{he} container technique has been increasingly adopted in operating system virtualization by cloud computing and edge computing due to its high convenience and flexibility in deploying diverse applications \cite{EDG}. The lightweight, scalable, and well-isolated environment provided by containers greatly increases the applications' portability and \linebreak availability \cite{PLD}, but at the cost of excessive communication overhead and resource utilization imbalance. Therefore, reducing communication overhead and balancing the usage of multi-type computing resources are critical design challenges for container-based multi-access edge computing (MEC).

Zhang et al. \cite{ZJZ} studied the joint task scheduling and containerization in an edge computing system, where the tasks associated with a given application were first scheduled into processors deployed on an edge server, then an appropriate algorithm was invoked to determine the containerization scheme. As a result, multiple containers, each representing a standard unit of software for packaging code and all related dependencies, were created for executing different tasks on individual processors, while considering the time cost of inter-container communications. Workflow is a common model for describing microservice-based application's tasks executed in containers. Bao et al. \cite{BLW} presented a microservice-based workflow scheduling algorithm for minimizing the end-to-end delay of a given single application consisting of multiple microservices under a pre-specified budget constraint in a cloud, where the inter-container communication latency was considered a component of the end-to-end delay. A dual-line container placement algorithm was proposed in \cite{RZZ} to decide the container placement positions in a container cluster while being aware of the inter-container communication traffic. We note that all of the above contributions concentrated on reducing communication overhead while ignoring computing resource utilization problems. Naturally, allocating the containers of the same application into the same server is beneficial for reducing the cost of inter-container communications.  However, this will cause serious resource utilization imbalance, since the same type of computing resource is usually used intensively by different containers of the same application, e.g., a CPU-intensive application \cite{LLV}.  Hence, without a proper resource-utilization balanced container scheduler, the particular server may suffer exponentially increasing response latency \cite{HYJ}, and the total throughput of the system may be significantly reduced. 

Therefore, it is important to also consider the balanced use of computing resources when designing scheduling algorithms.  Since each task has a different computing resource requirement, Li et al. \cite{LWL} considered the problem of scheduling microservice tasks of workflow applications to containers configured on on-demand virtual machines (VMs) and suggested a heuristic task scheduling algorithm for guaranteeing that the precedence-constrained or independent tasks are scheduled to available containers. As a benefit, the computing resources were more efficiently utilized. Ye et al. \cite{YLX} explored the stochastic hybrid workflow scheduling problem to keep the cost of computing resources to the minimum, by jointly scheduling offline and online workflows. Although these contributions considerably improve the utilization efficiency of computing resources, they impose high communication overhead, because the containers of the same application have to exchange control messages and payload data. Therefore, the effectiveness of communications among containers has a significant impact on how well services are provided.

How to keep computing resource utilization balanced while reducing inter-container communication costs remains a grave challenge.   In order to reduce the total inter-container communication traffic and increase the utilization efficiency of computing resources, \linebreak Wu et al. \cite{WZD} proposed a container placement strategy for containerized data centers, by exploiting the inter-container communication traffic pattern that obeys a Zipf-like distribution. 
%More specifically, their proposed method first groups containers into blocks based on the inter-container traffic correlation and then deploys container blocks that have distinct preferable resource types onto the same VM, without exceeding the threshold of the CPU utilization rate.  
%They considered the effects of container-level placement with respect to particular datasets.
They assumed that the major communication traffic originates from the containers that intensively communicate with each other. Each group of such containers is treated as an individual container block, which can be deployed either on the same server or on different servers, and the authors further assumed that the communication traffic between container blocks is negligible. Unfortunately, this assumption is unreasonable in the context of collaborative MEC servers.
In \cite{HU}, the computing resource requirements of containers are represented by a flow network, and container scheduling is formulated as a minimum-cost flow problem. To properly prioritize the execution of containers subject to a batch of concurrent requests, the proposed approach of  \cite{HU} considered the average life cycle of containers as well as the affinity between the containers and the servers. 
Their solution took container affinity into account and placed containers with affinity on the same server. However, it is improper for scheduling container clusters with high dependencies.
Lv et al. \cite{LLV} studied container placement and container reassignment strategies to balance resource use in large-scale Internet data centers, where the initial container distribution is optimized further by reassigning containers among servers. However, as far as a large number of distributed MEC servers are concerned, migrating containers among these servers may incur significant communication overhead.
%Their proposed scheme first obtains a set of container requests to build a streaming network, and then executes the successive shortest path algorithm to place the requested containers.
%They suggest using a successive shortest path algorithm to place the requested containers after first obtaining a set of container requests to build a streaming network. In contrast to them, we concentrate on container scheduling with dependencies.
%

Furthermore, the widely used container orchestration platforms, such as Docker, only provide simple rule-based scheduling principles.
Therefore, there has been a scarcity of efforts that focus on container scheduling algorithms capable of jointly reducing the communication overhead and balancing the computing resource usage in MEC. 

Against the above backdrop, in this paper, our novel contributions are summarized as follows. 
\begin{itemize}
	\item We propose a task containerization and container placement optimization framework for applications running on MEC servers from a joint communication and computing perspective. The proposed framework comprises two modules. The task containerization module jointly considers low inter-container communication overhead and balanced multi-type computing resource requirements of containers. The container placement module places instantiated containers to the appropriate MEC servers by considering the balanced usage of the multi-type computing resources on MEC servers. The proposed framework is capable of achieving both low communication overhead and balanced computing resource requirement among containers, as well as balanced computing resource utilization among servers. To the best of our knowledge, our work is the first to investigate the impact of task partitioning, task containerization, and container placement on inter-container communication overhead and resource utilization balancing.
	\item We offer a workflow modeling method for highly interdependent tasks of an application and propose a mathematical model of the workflow to reflect the interactions among the tasks, the communication overhead, and the computing resources needed by each task. Based on the workflow model of tasks, we present a method for calculating the communication overhead and the utilization efficiency deviation of multi-type computing resources in the MEC-based computing network considered. The proposed model and methods are directly applicable to various workflow-based cloud computing and edge computing platforms.
	\item We evaluate the proposed task-containerization -and-container-placement algorithms in multiple aspects through extensive experiments, and compare them with state-of-the-art methods. Our experiments show that the proposed methods are capable of reducing the communication overhead by up to 74.10\%, decreasing the normalized maximum load by up to 60.24\%, improving the CPU utilization efficiency by up to  30.66\%, and improving the memory utilization efficiency by up to 40.77\% under the considered system configurations.  
	%to demonstrate that it can produce high-quality allocation outcomes while requiring substantially less communication time and more equitable use of resources. Compared with the existing containerization scheme, our method can reduce 53.28\% of the transmission time, decrease the maximum normalized load by 60.24\% and improve 30.66\% of the resource utilization efficiency
\end{itemize}

%We divide the application tasks first. Grouping frequently communicated tasks into containers can minimize data transfer between different containers since inter-container communication is less effective than intra-container communication. The resource requirements of tasks must be factored in when splitting up applications in an effort to realize each container's needs balancing, making sure that the resource requirements of each group of tasks are as similar as possible.  Then, to balance the utilization of resources on the servers,  the container must be deployed on the appropriate server. We choose the optimum location for it depending on how closely the container's resource requirements match those of the server. Finally, using communication cost and resource balance cost as measurement parameters, we estimate the algorithm's performance and assess its efficiency. Compared with the existing containerization scheme, our method can reduce 53.28\% of the transmission time, decrease the maximum normalized load by 60.24\% and improve 30.66\% of the resource utilization efficiency when executing the application tasks on the edge computing system.

The rest of this paper is organized as follows. In Section %MDPI: we have revised the section label to regular number, please confirm.
\ref{sec2}, we present the MEC system model. In Section \ref{sec3}, we formulate the problem of jointly reducing the communication overhead and improving the degree of balance for multi-type computing resource utilization, which is solved by the algorithms proposed in Section \ref{sec4}, where the original problem is divided into two subproblems, namely the task containerization problem that is formulated as a graph partitioning problem \cite{TCG}, and the container placement optimization problem that is formulated as a {multi-choice vector bin packing problem \cite{PATTSHAMIR20121591}, which is a generalization of the classic vector bin packing problem}. In Section \ref{sec5}, the proposed algorithms are compared with state-of-the-art methods by extensive simulations, and our conclusions are drawn in Section \ref{sec6}.

%补充一个系统架构
%\section{system structure}

	\section{System Model}\label{sec2}
\begin{figure}[t]
	\includegraphics[width = 3.5in]{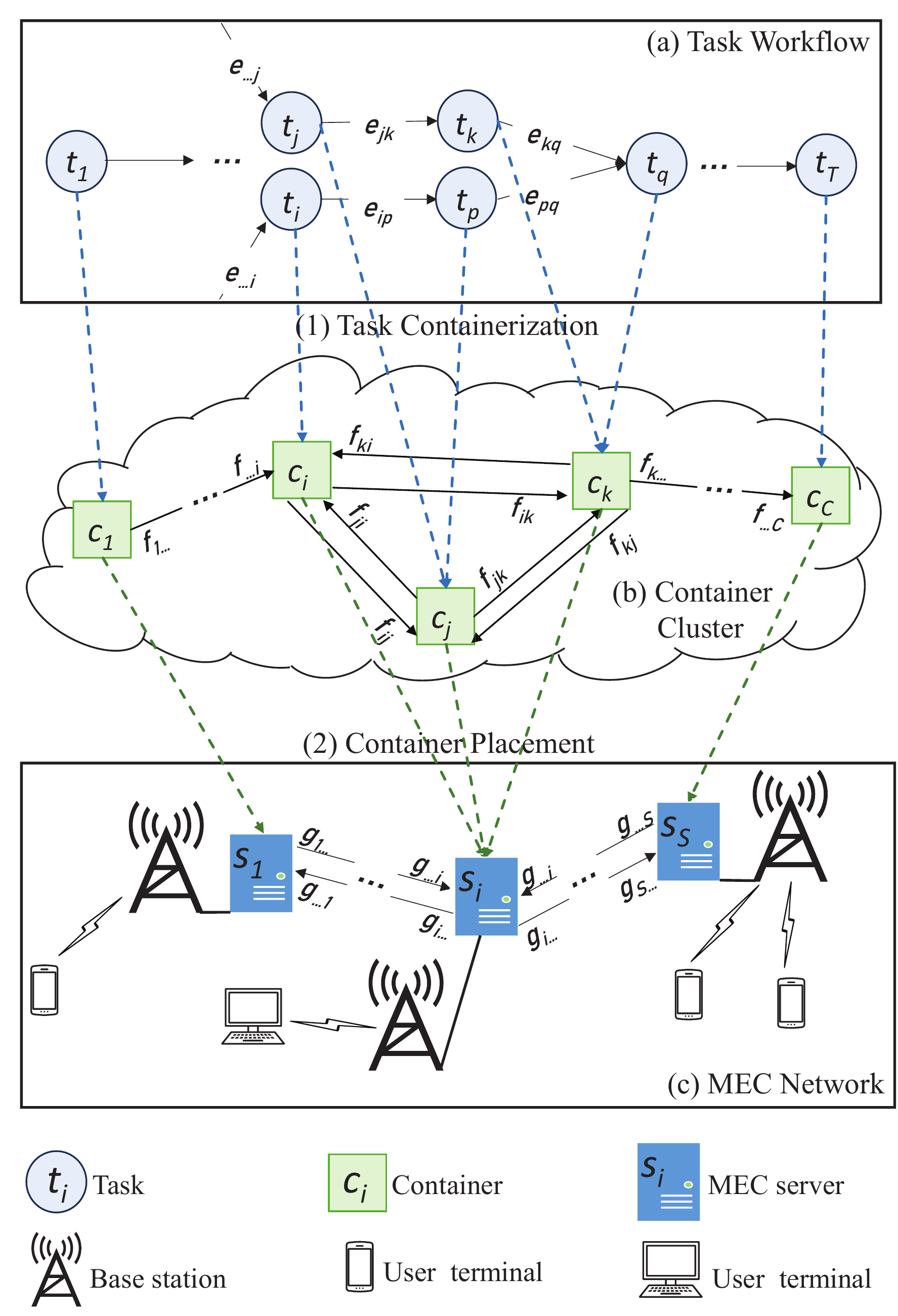}
	\caption{A system model for user terminals to access an MEC-based computing network.}
	\label{fig:arch}
\end{figure}

%In this section, we first describe the task containerized and container placement problem, then we go into the above trade-off by analyzing the overall costs and constraints.
%We suggest an offloading scenario for mobile edge computing that incorporates numerous edge computing servers, base stations, and terminal devices, as shown in Figure \ref{fig:arch}. The base stations are used by terminal devices to offload a sizable number of applications to the edge servers. The application is constructed as a set of collaboratively interconnected tasks. Edge computing systems containerize tasks and run them on edge servers.
%互联网服务提供商将应用程序放到边缘计算系统中部署和执行，每个应用程序可以实现特定的应用功能，满足特定的业务目标。如图所示，边缘计算服务器设置在基站附近，并与基站相连，终端设备可以通过基站接入边缘计算系统使用应用程序提供的功能。
%As shown in Figure \ref{fig:arch}, we propose an offloading scenario for mobile edge computing that includes numerous edge computing servers, base stations, and terminal devices. Terminal devices use base stations to offload a large number of applications to edge servers. The edge computing system containerizes applications made up of interconnected tasks and places them on edge servers to run.
As shown in Figure \ref{fig:arch}, we consider the use case where Internet service providers set up and execute delay-sensitive applications in a \textit{computing network} %MDPI: Please confirm if italic should be retained.
composed of multiple MEC servers.  The MEC servers are deployed at the edge of a mobile network and connected to base stations (BSs). The user terminals (UTs) can access diverse delay-sensitive computing-intensive services, such as cloud gaming and cloud virtual reality (VR), provided by the MEC-based computing network through BSs. All the MEC servers are distributively connected to each other by optic fibers, so that their computing power can be shared via cooperation.

The MEC servers are represented by $\mathcal{S} = \{ {s_1},{s_2}, \cdots ,{s_S}\}$, where $S = \left| \mathcal{S} \right|$ is the number of servers.
In the entire MEC service process, computing resources $\mathcal{R} = \{ {r_1},{r_2}, \cdots ,{r_R}\}$, such as CPU, memory and disk storage, are involved, where  $R = \left| \mathcal{R} \right|$ is the number of resource types. 
For server ${s_i} \in \mathcal{S}$, let {${\gamma}_{{s_i},{r_k}}$} denote the capacity of resource ${r_k} \in \mathcal{R}$.
We consider an application composed of multiple interdependent tasks $\mathcal{T} = \{ {t_1},{t_2}, \cdots ,{t_T}\}$, and the number of tasks is $T = \left| \mathcal{T} \right|$. Let the matrix %MDPI: Please confirm if all bold in Equation necessary. Answer: matrix and vector are represented by bold symbols.
$\mathbf{E} = {\left[ {{e_{ij}}} \right]_{T \times T}}$ represent dependencies between tasks, and we have 
\begin{equation}
	e_{ij} = \left\{
	\begin{array}{rcl}
		1, & \text{if task ${t_i}$ sends data to task ${t_j}$} \\
		0. & \text{otherwise}
	\end{array}
	\right.\label{eq:edge}
\end{equation}

The tasks are allocated to the container cluster $\mathcal{C} = \{ {c_1},{c_2}, \cdots ,{c_C}\}$, which comprises $C = \left| \mathcal{C} \right|$ containers. For container ${c_i} \in \mathcal{C}$, {${\beta}_{{c_i},{r_k}}$} denotes the demand for resource ${r_k} \in \mathcal{R}$. For each task ${t_i}$, let vector {${{\bf v}_i} = (v_i^{(r_1)},v_i^{(r_2)}, \cdots ,v_i^{(r_R)})$} represent {the individual volumes of various resources required %MDPI: Footnote is not allowed for this type of journal, we have moved it into maintext in highlighted blanket ( ), please confirm.
	by task $t_i$}. Note that in order to eliminate the difficulties encountered in evaluating the usage of heterogeneous resources, in this paper the volume of a specific type of resource required is normalized as the ratio of this partial volume to the total volume of the available resources of the same type across all the servers considered.  The matrix $\mathbf{D} = {\left[ {{d_{ij}}} \right]_{T \times C}}$ represents the tasks deployment, where we have  
\begin{equation}
	d_{ij} = \left\{
	\begin{array}{rcl}
		1, & \text{ if task ${t_i}$ is assigned to container ${c_j}$} \\
		0. & \text{otherwise}
	\end{array}
	\right.
\end{equation}

The resource ${r_k}$ demanded by the container ${c_j}$ is the sum of the resources required by the tasks assigned to ${c_j}$, namely
\begin{equation}
	%{\beta}({c_j},{r_k}) = \sum\limits_{{d_{ij}} \in \mathbb{D}} {{d_{ij}} \cdot v_i^k} 
	{{\beta}_{{c_j},{r_k}}} = \sum\limits_{i=1}^T {{d_{ij}} v_i^{({r_k})}}.
	%{\rm{U}}({c_j},{r_k}) = \sum\limits_{i \in T} {{d_{ij}} \cdot v_i^k} 
\end{equation}

Let the matrix $\mathbf{M} = {\left[ {{m_{ij}}} \right]_{C \times S}}$ denote the container placement strategy, as expressed by 
\begin{equation}
	m_{ij} = \left\{
	\begin{array}{rcl}
		1, & \text{if container ${c_i}$ is placed in server ${s_j}$} \\
		0. & \text{otherwise}
	\end{array}
	\right.
\end{equation}

\section{Problem Formulation}\label{sec3}
\subsection{Communication Overhead}
We consider the communication overhead incurred by the communications between containers within an MEC-based computing network.  %Communication cost is expressed in terms of the amount of data or number of communications between servers. 
%Especially, when the containers are located on different servers, communication time will be greatly increased due to the long distance between servers.
Typically, the life cycle of a task includes three stages: receiving data, running algorithms, and sending output data. For example, task ${t_i}$ first receives the data and stores it in memory, then it runs algorithms upon obtaining all the input information it needs. Finally, when the execution is finished, task ${t_i}$ sends the output data to related tasks, which takes a transmission time of ${\tau_{t_{i}}}$. This is also the communication time between containers, once the tasks are assigned to containers. 

Let $\mathbf{F} = {\left[ {{f_{ij}}} \right]_{C \times C}}$ denote the container dependency matrix. Upon assuming that the communication time between tasks assigned to {the same} container is negligible, the element of $\mathbf{F}$ can be expressed as:
\begin{equation}
	f_{ij} = \left\{
	\begin{array}{rcl}
		\sum\limits_q^T \sum\limits_p^T d_{pi}\tau_{t_p} e_{pq} d_{qj}, & i\ne j \\
		0. &i=j
	\end{array}
	\right.
\end{equation}
Similarly, %MDPI: the first letter is uppercase, please check if we need to add indent for this paragraph? if it is, please check the whole article. answer: no need to add indent.
we formulate the server dependency matrix as $\mathbf{G} = {\left[ {{g_{ij}}} \right]_{S \times S}}$, where we have 
\begin{equation}
	g_{ij} = \left\{
	\begin{array}{rcl}
		\sum\limits_{q}^C\sum\limits_{p}^C  m_{pi} f_{pq} m_{qj}, & i\ne j \\
		0. &i=j
	\end{array}
	\right.
\end{equation}
Upon assuming that the packet size and the number of packets required in each occurrence of communication remain constant, the communication overhead in the MEC-based computing network can be modeled as the ratio of the time spent in sending data between servers to the total communication time, namely
\begin{equation}
	{C_{\textrm {OH}}(\mathbf {D},\mathbf{M})} = \frac{{\sum\limits_{i}^S\sum\limits_{j}^S {g_{ij}}}}{\sum_{i=1}^{T} \tau_{t_{i}}}.
\end{equation}

\subsubsection{Computing Resource Utilization Balance}
The efficiency of servers is often degraded by the unbalanced use of the various computing resources deployed on the servers. The resource ${r_k}$ used by all the containers residing in the server ${s_i}$ is quantified by
\begin{equation}
	{\beta_{{s_i},{r_k}}(\mathbf {D},\mathbf{M})} = \sum_{j=1}^{C} {{m_{ji}}  } {\beta_{{c_j},{r_k}}}.
\end{equation}

Since the total available resource $r_k$ on the server $s_i$ is {$\gamma_{{s_i},{r_k}}$}.  Then, {$u_{{s_i},{r_k}}(\mathbf {D},\mathbf{M}) = \frac{{\beta_{{s_i},{r_k}}(\mathbf {D},\mathbf{M})}}{{{\gamma}_{{s_i},{r_k}}}}$} represents the utilization efficiency of resource  ${r_k}$ on server ${s_i}$. 
%We have chosen the utilisation rate $u({s_i},{r_b})$ of a resource ${r_b}$ as the criterion and the ratio of resource utilisation $u({s_i},{r_k})$ to $u({s_i},{r_k})$ reflects the utilisation difference between ${r_k}$ and ${r_b}$.
Typically, if a server's consumption of different types of computing resources is relatively balanced, the server is able to host more containers and more applications. Then,  the balance degree of multi-type computing resource utilization on server ${s_i}$ is given by 
\begin{equation}
	{b_{i}} = \sum_{k=1}^{R} \frac{{{( {u_{{s_i},{r_k}}(\mathbf {D},\mathbf{M}) - \bar u_{s_i}} )}^2}}{R},
\end{equation}
where {${\bar u_{s_i}}$} is the mean utilization efficiency of all resources on server $s_i$. 

This metric reflects the deviation of the utilization efficiency of all resource types on a server from the average resource utilization efficiency.
The total resource balance degree of the MEC-based computing network is the sum of the resource balance degree of all servers, and it is defined as
\begin{equation}
	{{B_{\textrm {tot}}(\mathbf {D},\mathbf{M})}} = \sum_{i=1}^{S} {b_{i}}.\label{total_deviation}
\end{equation}

Therefore, the joint optimization of the communication overhead and the degree of resource utilization balance is formulated as a weighted-sum minimization problem of %MDPI: Please confirm if sub-equations are necessary. Answer: Yes, they are necessary. 
%\vspace{-6pt}
% \begin{equation}
	% \begin{aligned}
		% &{\textrm {min}} ~\mu_{C} C_{\textrm {OH}}+ \mu_{B}B_{\textrm {tot}} \\
		% &{\textrm {s.t.}}~{\beta}({s_i},{r_k}) \le {\gamma}({s_i},{r_k}),\quad \forall {s_i} \in {\cal S},\forall {r_k} \in {\cal R}. \end{aligned} \label{problem_1}
	%  \end{equation}
\begin{subequations}\label{problem_1}
	\begin{align}
		\mathsf{P}1: {\mathop{\min}_{\mathbf{D}, \mathbf{M}}} \quad &\mu_{C} C_{\textrm {OH}}(\mathbf {D},\mathbf{M})+ \mu_{B}B_{\textrm {tot}}(\mathbf {D},\mathbf{M}) \label{P1_obj} \\
		{\textrm {s.t.}}\quad & \mathbf{D}\in\{0,1\}^{T\times C}, \label{constraint_1_P1} \\
		& \mathbf{M}\in \{0,1\}^{C\times S}, \label{constraint_2_P1}\\ 
		& \sum_{j=1}^C d_{ij} = 1,\label{constraint_3_P1}\\
		& \sum_{j=1}^S m_{ij} = 1,\label{constraint_4_P1} \\
		& 0\le{\beta}_{{s_i},{r_k}}(\mathbf {D},\mathbf{M}) \le {\gamma}_{{s_i},{r_k}}.\label{constraint_5_P1}
	\end{align}
\end{subequations}

In the above optimization problem,  $\tau_{t_{i}}\ge 0$, $v_t^{(r_k)}\ge 0$, $0\le \mu _{C} \le 1$, $0\le \mu _{B} \le 1$, ${t_i} \in {\cal T}$, ${s_i} \in {\cal S}$, ${r_k} \in {\cal R}$, and $ {c_j} \in {\cal C}$ are all constants known a priori. Here, $\tau_{t_{i}}\ge 0$ indicates that the transmission time of each task is non-negative; $v_t^{(r_k)}\ge 0$ indicates that each task has a non-negative demand for each type of resource; while $0\le \mu _{C} \le 1$ and $0\le \mu _{B} \le 1$ specify the value range of the weights corresponding to each of the component objectives.  {\eqref{constraint_1_P1} and \eqref{constraint_2_P1} indicate that the optimization variables are matrices composed of binary-value elements. Equations \eqref{constraint_3_P1} and \eqref{constraint_4_P1} imply that each given task $t_i$ and given container $c_i$ have  to be allocated to a single container and a single server, respectively. Furthermore, \eqref{constraint_5_P1} represents that the resource $r_k$ used by all the containers residing in the server $s_i$ must not exceed the capacity of the resource $r_k$ on the server $s_i$. }

%The specific form of the minimization problem is given by
%其中，第一行约束表示每个任务的传输时间非负，第二行约束表示每个任务对每种种类资源的需求非负，第三行约束表示加权参数的取值范围，第四行约束表示对服务器的需求少于服务器具有的资源容量。最小化问题的具体形式为：
%将最优化问题展开

% \begin{equation}
	% \begin{aligned}
		%&\quad \mu_{C} C_{\textrm {OH}}+ \mu_{B}B_{\textrm {tot}}\\
		% &=\mu _{C} \frac{\sum g_{i j}}{\sum_{i=1}^{T} \tau_{t_{i}}}+\mu _{B} \sum_{i=1}^{S}\sum_{k=1}^{R} \frac{(\frac{\beta({s_i},{r_k})}{\gamma(s_{i}, r_{k})}-\bar{u}(r))^{2}}{R}\\
		% &=\mu _{C} \frac{\sum g_{i j}}{\sum_{i=1}^{T} \tau_{t_{i}}}+\mu _{B} \sum_{i=1}^{S}\sum_{k=1}^{R} \frac{(\frac{\sum_{j=1}^{C} m_{ji} \beta(c_{j}, r_{k})}{\gamma(s_{i}, r_{k})}-\bar{u}(r))^{2}}{R}\\
		% &=\mu _{C} \frac{\sum g_{i j}}{\sum_{i=1}^{T} \tau_{t_{i}}}+\mu _{B} \sum_{i=1}^{S}\sum_{k=1}^{R} \frac{(\frac{\sum_{j=1}^{C} m_{ji} \sum\limits_{t=1}^T {{d_{tj}} v_t^{(r_k)}}}{\gamma(s_{i}, r_{k})}-\bar{u}(r))^{2}}{R}\\
		%\end{aligned} \label{problem_01}
		%  \end{equation}
	
	%使通信开销变小倾向于将任务划分到一个容器里且容器放置到一个服务器上，而这也必然会导致服务器集群中计算资源利用率的变小。与其在一个容器或服务器里装尽可能多的东西，我们可以试着平衡通信开销和计算资源利用率。
	{It is worth noting that reducing communication overhead may lead to putting more tasks into fewer containers and packing as many containers as possible into fewer servers. However, this may result in a lower utilization efficiency of computing resources in the servers. In other words, there is a trade-off between the communication overhead ${C_{\textrm{OH}}}$ and the computing resource utilization efficiency ${B_{\textrm{tot}}}$. 
		We attempt to balance ${C_{\textrm{OH}}}$ and ${B_{\textrm{tot}}}$ rather than cram as many items as possible in a container or a server.}
	%可以看出，本联合优化问题中C和B两项相互影响。为了解决此问题，我们依据应用程序部署到服务器上的两个阶段，即任务分组容器化和容器放置到服务器上，将整体问题拆分为两个问题，在任务分组容器化时目标为容器之间的通信开销最小和每个容器所需计算资源相似。
	
	The problem $\mathsf{P}1$ is essentially a multi-objective combinatorial optimization problem involving two subproblems, namely graph partitioning \cite{TCG} that is corresponding to task containerization  and multi-choice vector bin packing \cite{PATTSHAMIR20121591} that is corresponding to container placement, as shown in Figure~\ref{fig:arch}. Both of them are known NP-hard problems. Hence, the existence of any polynomial time optimal solution is ruled out unless ${\mathsf P} = \mathsf {NP}$. Furthermore, as mentioned before, the component objectives ${C_{\textrm{OH}}}$ and ${B_{\textrm{tot}}}$ may conflict with each other, and both of them rely on the optimization variables $\mathbf D$ and $\mathbf M$. Theoretically, the optimal solution to $\mathsf{P}1$ can be found by brute-force search. However, in practice, there may be a large number of tasks, containers, and servers to deal with. Hence, the computational complexity of obtaining the optimal solution to the joint optimization problem  $\mathsf{P}1$ is prohibitive. Alternatively, it is possible to solve $\mathsf{P}1$ more efficiently by using an iterative method, where at each iteration, the objective of  $\mathsf{P}1$ is minimized with respect to one of the two matrix variables $\mathbf D$ and $\mathbf M$ while the other matrix variable is held fixed. Nevertheless, the deployment of this strategy may require a dedicated centralized computing server to run the scheduling algorithms and is not naturally aligned with the physically sequential processes of an MEC-based computing network, where the task containerization takes place first and is then followed by container placement. In other words, in practice, it is meaningful to first obtain $\mathbf D$  and then obtain $\mathbf M$ relying on the result of $\mathbf D$. 
	
	With this in mind, we turn our attention to a degenerate version of $\mathsf{P}1$ so that the reformulated problem is more naturally matched to the practical sequential process of deploying an application to servers. As a result, a non-iterative sequential strategy can be invoked, which is expected to find good solutions capable of optimizing both the inter-container communication overhead and the degree of resource utilization balance at the expense of lower computational complexity. More specifically, the goal of task containerization is to minimize the weighted-sum of the inter-container communication overhead and the average deviation of multi-type resource demands by each container. Thus, we have
	%\vspace{-6pt}
	\begin{subequations}\label{problem_2}
		\begin{align}
			\mathsf{P}2: {\mathop{\min}_{\mathbf{D}}} \quad &\mu _{C}\frac{{\sum\limits_{i=1}^C\sum\limits_{j=1}^C {{f_{ij}}} }}{{\sum\limits_{i = 1}^T {{\tau _{{t_i}}}} }} + {\mu _B}\sum\limits_{k = 1}^R {\sum\limits_{j = 1}^C {\frac{{{{(\beta _{{c_j},{r_k}} - \overline \beta_{r_k} )}^2}}}{C}} } \label{P2_obj}\\
			{\textrm {s.t.}}\quad & \mathbf{D}\in\{0,1\}^{T\times C}, \label{constraint_1_P2} \\
			& \sum_{j=1}^C d_{ij} = 1,\label{constraint_3_P2}
		\end{align} 
	\end{subequations}
	%划分任务为多个小组并容器化本质上属于图划分问题，然而与经典的图分割问题相比，我们在求解割集时，不仅考虑边的权重即通信开销，还考虑了顶点的权重即资源需求。
	where $ \overline \beta_{r_k} $ represents the average value of all containers' demand for resource $r_k$. As pointed out before, dividing tasks into groups and putting each group of tasks into a container can essentially be formulated as a graph partitioning problem \cite{TCG}. Different from the traditional treatment of graph partitioning problems, in this paper we not only take into account the weights of edges, which represent the communication overhead but also consider the weights of vertices, which represent the resource requirements of a particular task.
	%在任务容器化求出D的情况下
	
	Furthermore, we simplify the container placement problem without considering the communication overhead caused by inter-container communications and the communication time  between servers in the objective function. This is because the servers' geographic locations have a significant impact on the communication overhead between servers and it can be difficult to ascertain the servers' locations. After the task containerization module obtains $\mathbf{D}$, the container placement module is intended to minimize the deviation of the utilization efficiency of various computing resources on the MEC servers, hence the container placement problem is formulated as:
	\begin{subequations}\label{problem_3}
		\begin{align}
			\mathsf{P}3: {\mathop{\min}_{\mathbf{M}}} \quad &{\mu _B}\sum_{i = 1}^S\sum_{k=1}^{R}\frac{(u_{{s_i},{r_k}}(\mathbf{M}) - \bar u_{s_i})^2}{R} \label{P3_obj}\\ 
			{\textrm {s.t.}}\quad & \mathbf{M}\in \{0,1\}^{C\times S}, \label{constraint_3_P3}\\
			&\sum_{j=1}^S m_{ij} = 1,\label{constraint_3_P3}\\
			&0\le u_{{s_i},{r_k}}(\mathbf{M}) \le1.\label{constraint_5_P3}
		\end{align}
	\end{subequations}
	%让每个容器所需要的计算资源
	%容器放置到服务器上的目标是最小化MEC服务器上各种类的计算资源利用率偏差。
	%容器放置问题与多维向量装箱问题相似，与经典多维向量装箱问题不同的是，我们给定一定数量的服务器，每个服务器具有特定的多种资源容量，我们的向量集里面的元素是从任务容器化阶段得到的容器，容器具有不同种类的计算资源需求，因此容器取值不是0或1，而是多种类资源需求向量。我们的目标也更为复杂，不是最小化服务器（“箱子”）的数量，而是将容器放置到服务器集群中需要让容器所需资源与服务器能提供的资源匹配。
	
	The container placement problem is similar to the multi-choice vector bin packing problem \cite{PATTSHAMIR20121591}. In contrast to the traditional multidimensional vector bin packing problem, we have a set number of servers and a range of resource capacities for each server. The components of our vector set come from the containers obtained during the task containerization stage. Container values are not 0 or 1, but rather a vector of distinct forms of resource requirements as containers have different types of computing resource requirements. Our objective is more complicated---instead of minimizing the number of servers, placing containers into MEC servers requires matching the resources the containers need with what the servers can provide.
	%具体的求解方法将在第三章说明。
	
	On the basis of making references to classic problems, it is vital to create solutions that are more fitting for our objectives. The fourth chapter will provide detailed explanations of the specific approach.
	
	\section{Task Containerization and Container Placement Schemes}\label{sec4}
	%缺少启发式算法的说明，NP-hard问题的说明，缺少初始调度的说明
	%First, we will discuss our optimization goals. Then, we will introduce the solution to the optimization problem. Based on the above discussions, we formally define the problems to be addressed in this paper.
	As shown in Figure \ref{fig:arch}, the entire process of performing edge computing in an MEC-based computing network can be sequentially divided into two stages: first allocating tasks to containers, and then placing containers onto servers. To find a high-accuracy approximate solution to the problem \eqref{problem_1}, we can consider the task containerization and the container placement sequentially. 
	\subsection{Task Containerization Method}
	A task containerization method is invoked to allocate tasks to containers. First of all, it is necessary to balance all types of resources needed by each container for achieving an equalized consumption of resources on each server.
	To this end, it is preferable to avoid grouping tasks that have a high demand for the same type of resource together. On the other hand, it is recommended to put tasks that have a large amount of data transmitted between each other into the same container and then place the containers having frequent communications with each other on the same server, in order to reduce the communication overhead.
	%Hence the design goal of the task division method is:
	%考虑是否要放此公式
	%\begin{equation}
	%\label{deqn_ex1a}
	%    \min \sum\limits_{{c_i},{c_j} \in C,i \ne j} {g({c_i},{c_j})} +\sum_{c_i  \in C} \sum_{r_k \in R}
	%    \frac{{\lvert U(c_i,r_k)-\overline U(c_i) \rvert}^2}{R}
	%\end{equation}
	
\begin{figure}[t]
		\includegraphics[width = 3 in]{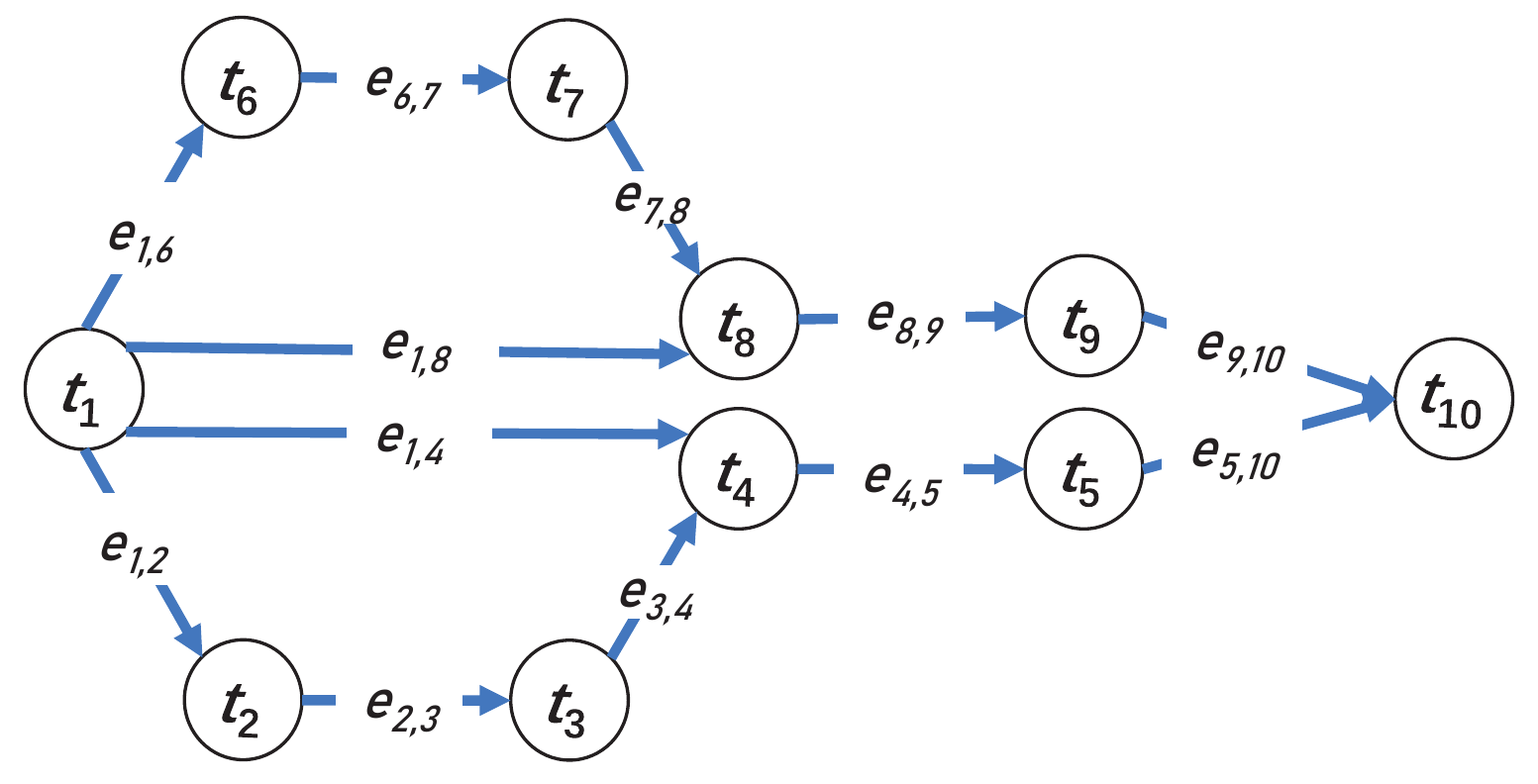}
		\caption{An example of DAG workflow with ten tasks.}
		\label{fig:workflow}
	\end{figure}

	%Workflows are a common model for describing applications executed in edge computing. 
	We use workflow to describe the relationship between different tasks of an application. Workflow is an important concept for describing cloud computing and edge computing applications. It consists of multiple interdependent tasks which are bound together through data or functional dependencies. As shown in {Figure~\ref{fig:workflow}}, an application can be built as a workflow, where each task implements certain functionality and collaborates with the others. In Figure~\ref{fig:workflow} the workflow is modelled, as a directed acyclic graph (DAG) $\mathcal{G} = (\mathcal{T},\mathcal{E})$. Each vertex ${t_i} \in \mathcal{T}$ represents a task, and the weight vector of vertex $t_i$ is $\mathbf{v}_i$, which represents various resources required by ${t_i}$. The {adjacency} matrix of the DAG is $\mathbf E$ and its elements are defined by \eqref{eq:edge}{, where} the weights of edges are not considered. The number of edges is denoted as $E$, i.e., $\left|{\mathcal{E}} \right|=E$, which equals the number of nonzero elements in $\mathbf E$. Furthermore, let ${\omega ({e_{ij}})}$ denote the weight of ${e_{ij}}$ and it represents the length of communication time from task $i$ to task $j$. Then, the elements $e_{ij}$ of $\mathbf E$ are updated as ${\omega ({e_{ij}})}$.  
	
	For a cluster of vertices $\mathcal {A} \subseteq \mathcal{T}$, let $\mathcal E(\mathcal A, \mathcal{A})$ denote the set of edges with all vertices in $\mathcal A$, and let ${\mathcal E}({\mathcal A},\mathcal{T}\backslash \mathcal A)$ represent the collection of edges {connecting the two disjoint vertex sets of} the graph cut $({\mathcal A},\mathcal{T}\backslash \mathcal A)$. 
	%All the vertices are partitioned into $L$ sets, and each vertex is only located in one set.
	Furthermore, define a vertex set $\mathcal{P} = ({{{\mathcal A}_1},{{\mathcal A}_2}, \cdots ,{{\mathcal A}_C}})$ as multiple disjoint vertex sets whose union is $\mathcal{T}$, i.e., we have
	%$\forall {A_i},{A_j} \in P,i \ne j,{A_i} \cap {A_j} = \phi$,
	${\bigcup\nolimits_{i = 1}^C {\mathcal A} _i} = \mathcal{T}$ and ${{\mathcal A}_i} \cap {{\mathcal A}_j} = \emptyset$, $\forall i \ne j$, $C \ge 1$. {Furthermore, we} denote with ${v_{{\mathcal A}_i}^{(k)}}$ the demand of resource ${r_k}$ by the subset ${{\mathcal A}_i}$.

	%图划分问题已被证明是NP难问题，启发式算法通常被使用来解决此类问题。
	%在本文中我们基于启发式算法的思想，先确定分区个数，每个分区内放入一个顶点做初始划分，对于剩下的顶点，设置一个打分函数，打分函数综合考虑了顶点权重和边权重，每个顶点根据分数选择分区，确保顶点的放置可以使分区内顶点权重不过载，并且分区之间边权重较小。
	%引用文献  
	It has been established that the graph partitioning problem is NP-hard. For finding approximate solutions to the graph partitioning problem considered more efficiently, we propose two heuristic algorithms in this paper, i.e., the partitioning with non-critical path based initialization {(P-NCPI)} and the partitioning with random initialization (P-RI). The major operations of the proposed partitioning algorithms include determining the number of disjoint vertex sets, assigning one vertex to each of the disjoint vertex sets in the initial partition, and setting a scoring function for the remaining vertices. Both vertex weights and edge weights are taken into account in the scoring function, and each vertex is exclusively assigned to one of the disjoint vertex sets based on its score. Typically, these operations can be implemented in an appropriate container orchestration software, such as Kubernetes or KubeEdge. This process guarantees that the vertices are partitioned so that the edge weights between the disjoint vertex sets are minimal and that the vertex weights within the disjoint vertex sets are not excessive.

	The {NCPI} and RI algorithms employed in our initial partition are described as follows.
	\begin{itemize}
		\item {NCPI}: Tasks are sorted topologically, and the critical path is chosen as the one with the highest weight. The tasks that comprise the critical path are known as essential tasks, and they determine the minimum completion time of an application. Due to the high communication overhead between essential tasks, we first select the required number of non-essential tasks and then assign each non-essential task to a different disjoint vertex set as an initial partition to reduce communication overhead between sets. 
		\item RI: Selecting the desired tasks randomly according to the number of disjoint vertex sets and assigning each of them to a different disjoint vertex set as an initial partition.
	\end{itemize}

	%Then calculate $\delta h(t,{A_i})$ of the remaining tasks to determine which partition the tasks are put into.
	
	To more conveniently describe the process of grouping the remaining vertices, we consider an alternative objective function with two components: the cost of inter-partition edge weights $C_e$ and the cost of intra-partition vertex weights $C_v$. For each given set of vertices $\mathcal{P} = ({{{\mathcal A}_1},{{\mathcal A}_2}, \cdots ,{{\mathcal A}_C}})$ and a specific resource ${r_k}$, this objective function is defined as:
	\begin{equation}
		\begin{aligned}
			f(\mathcal{P}) &= \mu_{C}{C_e}(\Omega (\mathcal{E}({{\mathcal A}_1},\mathcal{T}\backslash {{\mathcal A}_1})), \cdots ,\Omega (\mathcal{E}({{\mathcal A}_C},\mathcal{T}\backslash {{\mathcal A}_C}))) \\
			&+ \mu_{B}{C_v}(v^{(k)}_{{\mathcal A}_1}, \cdots ,v^{(k)}_{{\mathcal A}_C}),
		\end{aligned}
		\label{fp}
	\end{equation}
	where $\Omega (\mathcal{E}({{\mathcal A}_i},\mathcal{T}\backslash {{\mathcal A}_i}))$ represents the sum of weights corresponding to the edges contained in the set $\mathcal{E}({{\mathcal A}_i},\mathcal{T}\backslash {{\mathcal A}_i})$, and we have
	\begin{equation}
		\begin{aligned}
		&{C_e}(\Omega (\mathcal{E}({{\mathcal A}_1}, \mathcal{T}\backslash {{\mathcal A}_1})), \cdots,\Omega (\mathcal{E}({{\mathcal A}_C}, \mathcal{T}\backslash {{\mathcal A}_C}))) \\
		&= \sum\nolimits_{i = 1}^C {{\Omega} (\mathcal{E}}({{\mathcal A}_i},{\mathcal{T}}\backslash {{\mathcal A}_i})),\label{C_e}
		\end{aligned}
	\end{equation}		
	\begin{equation}
		{C_v}(v ^{(k)}_{{\mathcal A}_1}, \cdots ,v^{(k)}_{{\mathcal A}_C})= \sum\nolimits_{i = 1}^{C} g(v^{(k)}_{{\mathcal A}_i}),\label{C_v}
	\end{equation} 
	with ${g(v^{(k)}_{{\mathcal A}_i})}$ representing an increasing convex function of the vertex weights in set ${{\mathcal A}_i}$, and ${C_v}(\cdot)$ being a linear function for balancing the cost of vertex weights across different sets ${\mathcal A}_i$, $i=1, 2, \cdots, C$.
	
	It should be noted that the optimization variable in the objective function \eqref{P2_obj} is $\mathbf{D}$, while that in \eqref{fp} is $\mathcal{P}$. Both $\mathbf{D}$ and $\mathcal{P}$ represent the mapping of tasks to containers, and both the expression ${\sum\nolimits_{i=1}^C\sum\nolimits_{j=1}^C {f_{ij}} }$ in \eqref{P2_obj}  and the expression ${C_e}(\cdot)$ in \eqref{fp} characterize the communication time between containers. However, the differences are: 1) ${\sum\nolimits_{i=1}^C\sum\nolimits_{j=1}^C {f_{ij}} }$ in \eqref{P2_obj} is further normalized by dividing the total communication time of all the tasks, while ${C_e}(\cdot)$ is a direct summation of  $\Omega (\mathcal{E}({{\mathcal A}_i},\mathcal{T}\backslash {{\mathcal A}_i}))$, $i=1, 2, \cdots, C$;  2) $\sum\nolimits_{k = 1}^R {\sum\nolimits_{j = 1}^C {\frac{1}{C}{{{(\beta _{{c_j},{r_k}} - \overline \beta_{r_k} )}^2}}}}$ in \eqref{P2_obj} is the average deviation of multi-type resource demands by each container, while ${C_v}(\cdot)$ in \eqref{fp} is the total demand for computing resources by all the containers. Obviously, when the demands for multi-type computing resources by the individual containers are equal, the average deviation of multi-type resource demands by each container reaches the minimum. Therefore, \eqref{P2_obj} and (\ref{fp}) can be seen as different mathematical models for describing the same objective. Compared with \eqref{P2_obj}, Equation (\ref{fp}) directly expresses the communication time and computing resource requirements as specific analytical expressions of the variables ${{\mathcal A}_i}$, $i=1, 2, \cdots, C$, thus making it more convenient to derive the solution algorithm from the graph partitioning perspective.
	
	According to Equation (\ref{fp}), the task containerization method can be derived by solving the following optimization problem:
	\begin{equation}
		\begin{aligned}
			\mathsf{P}4: {\min_{\mathcal{P} = ({{\mathcal A}_1}, \cdots ,{{\mathcal A}_C})}}  f(\mathcal{P}) &= \sum\nolimits_{i = 1}^C {\mu_{C}{\Omega} (\mathcal{E}({{\mathcal A}_i},{\mathcal{T}}\backslash {{\mathcal A}_i}))}  \\ 
			&+ \sum\nolimits_{i = 1}^C {\mu_{B}g(v ^{(k)}_{{\mathcal A}_i}) }\\
			{\textrm {s.t.}}\quad & {\bigcup\nolimits_{i = 1}^C {\mathcal A} _i} = \mathcal{T},\\
			&{{\mathcal A}_i} \cap {{\mathcal A}_j} = \emptyset, \forall i \ne j.
		\end{aligned}
	\end{equation}

	For the convex function ${g(x)}$, based on \cite{TCG}, we select the family of functions $c(x) = \alpha {x^\theta }$, where $\alpha  > 0$ and $\theta  \ge 1$. The parameter $\theta$ controls the balance degree of vertex weights between partitioned sets. The greater the value of $\theta$, the greater the cost of imbalanced set weights. $\theta =1$ means that the imbalance between the demand of resource $r_k$ by the set ${{\mathcal A}_i}$, i.e.,  $v^{(k)}_{{\mathcal A}_i}$, is ignored. We choose $\alpha  = E\frac{C^{\theta  - 1}}{T^\theta }$ to serve as a proper scaling factor. The optimization problem $\mathsf{P}4$ is then written as:
	\begin{equation}
		\begin{aligned}
			\mathsf{P}5: {\min _{\mathcal{P} = ({{\mathcal A}_1}, \cdots ,{{\mathcal A}_C})}}f(\mathcal{P}) &= \frac{{\sum\nolimits_{i = 1}^C {\mu_{C}{\Omega} (\mathcal{E}({{\mathcal A}_i},\mathcal{T}\backslash {{\mathcal A}_i}))} }}{E} \\
			&+ \frac{1}{C}\sum\nolimits_{i = 1}^C {\mu_{B}{{\left(\frac{{v^{(k)}_{{\mathcal A}_i}}}{{\frac{T}{C}}}\right)}^\theta }}\\
			{\textrm {s.t.}}\quad & {\bigcup\nolimits_{i = 1}^C {\mathcal A} _i} = \mathcal{T},\\
			&{{\mathcal A}_i} \cap {{\mathcal A}_j} = \emptyset, \forall i \ne j.
		\end{aligned}
		\label{Pop}
	\end{equation}

	Furthermore, we define a function ${h}$ as:
	\begin{equation}
		\begin{aligned}
			h(\mathcal{P}) &= \sum\nolimits_{i = 1}^C \left({\mu_{C}{\Omega} (\mathcal{E}({{\mathcal A}_i},{{\mathcal A}_i})) - } \mu_{B}g(v ^{(k)}_{{\mathcal A}_i}) \right) \\
			&= \sum\nolimits_{i = 1}^C \left( {\mu_{C}{\Omega} (\mathcal{E}(\mathcal{T},\mathcal{T}) - \mathcal{E}({{\mathcal A}_i},\mathcal{T}\backslash {{\mathcal A}_i})) - } \mu_{B}g(v ^{(k)}_{{\mathcal A}_i})\right) \\
			%&= \sum\nolimits_{i = 1}^l {\omega (e(T,T)) - \omega (e({A_i},T\backslash {A_i})) - } c({v ^k}({A_i})) \\
			%&= l \cdot \omega (E) - (\sum\nolimits_{i = 1}^l {\omega (e({A_i},V\backslash {A_i})) + } c({v ^k}({A_i}))) \\ 
			&= C \cdot \mu_{C}{\Omega} (\mathcal{E}(\mathcal{T},\mathcal{T})- f(\mathcal{P}).
		\end{aligned}
		\label{hp}
	\end{equation}

	Therefore, the problem of minimizing ${f(\mathcal{P})}$ can be solved by maximizing ${h(\mathcal{P})}$. For any $1 \leq i < j \leq C$, if $h({{\mathcal A}_1}, \cdots ,{{\mathcal A}_i} \cup \{ t\} , \cdots ,{{\mathcal A}_j}\backslash \{t\}, \cdots ,{{\mathcal A}_C}) \ge h({{\mathcal A}_1}, \cdots ,{{\mathcal A}_i}\backslash\{t\}, \cdots ,{{\mathcal A}_j} \notag \\ \cup \{ t\} , \cdots ,{{\mathcal A}_C})$, vertex $t$ is assigned to {vertex set} ${A_i}$. Based on this condition, we define a difference function
	\begin{equation}\label{eq18}
		\begin{aligned}
			\Delta h(t,{{\mathcal A}_i}) = & h({{\mathcal A}_1}, \cdots ,{{\mathcal A}_i} \cup \{ t\} , \cdots ,{{\mathcal A}_j}\backslash \{t\}, \cdots ,{{\mathcal A}_C}) \\
			&- h({{\mathcal A}_1}, \cdots ,{{\mathcal A}_i}, \cdots ,{{\mathcal A}_j}, \cdots ,{{\mathcal A}_C}),
		\end{aligned}
	\end{equation}
	and assign vertex $t$ to vertex set ${\mathcal A}_i$ when we have $\Delta h(t,{{\mathcal A}_i}) \ge \Delta h(t,{{\mathcal A}_j})$. 
	Let $N(t)$ denote the neighbors of vertex $t$ (i.e., the vertices directly connected to vertex $t$). According to Equation (\ref{hp}), the difference function is rewritten as:
	\begin{equation}
		\begin{aligned}
			\Delta h(t,{{\mathcal A}_i})  = & \mu_{C}\Omega (\mathcal{E}(N(t) \cap {{\mathcal A}_i}, N(t) \cap {{\mathcal A}_i}))\\
			&  - \mu_{C}\Omega (\mathcal{E}(N(t) \cap {{\mathcal A}_j}, N(t) \cap {{\mathcal A}_j}))  \\
			& + \mu_{B}\left( g(v ^{(k)}_{{\mathcal A}_i}) + g(v ^{(k)}_{{\mathcal A}_j})\right)\\
			& - \mu_{B}\left(g(v ^{(k)}_{{\mathcal A}_i \cup \{t\}}) + g(v ^{(k)}_{{\mathcal A}_j \backslash \{t\}})\right).
		\end{aligned}
		\label{inter}
	\end{equation}
	
	The details of the proposed task containerization method are shown in Algorithm~\ref{alg:tca}. First, tasks are selected by the RI or NCPI algorithm for initial partitioning (Lines 2--35 of Algorithm~\ref{alg:tca}).  Once the initial partitioning has been determined, the remaining tasks are partitioned using the scores derived from the above calculations (Lines 36--47 of Algorithm~\ref{alg:tca}), thus ensuring that the communication cost and the multi-type computing resource balancing cost are minimized during the task containerization stage.
	%Once the initial partition has been determined, the remaining nodes are grouped using the scores derived from the above calculations, thus ensuring that communication costs and resource balancing costs are minimized during the containerization phase of the task.
	
	In order to measure whether the resources required for each vertex set are balanced, we define the measurement parameter $\lambda$ as the normalized maximum load:
	\begin{equation}
		\lambda  = C\cdot \frac{\mathop {\max}_j \{\sum_{k=1}^{R} {{\beta}_{{c_j},{r_k}}|{c_j}\in\mathcal{C} \}}} {{\sum_{j=1}^{C} {\sum_{k=1}^{R} {{\beta}_{{c_j},{r_k}}}}}}.\label{normalized_maximum_load}
	\end{equation}
	The balance degree of the usage of multi-type computing resources improves as $\lambda$ approaches 1.
	
	\subsection{Container Placement Method}
	%当任务划分成多个分区后。每个分区就是一个容器，本节研究容器的放置问题。
	%容器放置到服务器上主要考虑两点：一是容器所需要的资源和服务器能提供的资源的匹配程度，二是考虑到服务器之间的距离，需要将通信频繁和传输量大的容器放置到同一个或相邻的服务器上。由于边缘节点的位置一般都相距较远，且距离对上层应用难以获得，我们选择将这样的容器放到同一服务器上。
	When tasks are divided into multiple vertex sets, each set is a container, and the resources are available to each container. We conduct a comparative study of container placement methods by using the dot product (DP) algorithm \cite{PRT} and the first fit decreasing (FFD) algorithm \cite{PRT} to individually solve the problem $\mathcal{P}3$.
	%This section shows the placement method of containers.
	%There are two main considerations when placing containers on servers: one is the matching degree between the resources required by each container and the resources provided by each server; the other is that considering the distance between servers, containers with frequent communication and large transmission volume need to be placed on the same or adjacent servers. As the location of edge nodes is generally far away, and the distance is difficult to obtain for upper--layer applications, we choose to put such containers on the same server.

	%We get the match between each container and each server using the bin packing algorithm and place the container on the server with the highest match.
	
	DP considers the demands of multi-type computing resources by each container as well as the capacities of multi-type resources on each server. Any server is listed as a candidate server for a container if it has enough multi-type resources to accommodate the container. For each container ${c_i}$ and one of its candidate servers ${s_j}$, the dot product of multi-type resources required by the container and multi-type resources that the server can provide is denoted as $\textsf{DP}_{ij}$. 
	The larger the value of $\textsf{DP}_{ij}$, the more resources server ${s_j}$ is able to supply to container ${c_i}$. To improve computing resource utilization, we tend to place the container on the server that has the greatest amount of resources, hence we use $\textsf{DP}_{ij}$ as the matching degree between demand and supply, and for servers that are not candidates, we set the matching degree as 0. Thus we have 
	\begin{equation}
		\textsf{DP}_{ij} = \left\{
		\begin{array}{rcl}
			\sum_{k=1}^{R} {{\beta}_{{c_i},{r_k}} \cdot } {\gamma}_{{s_j},{r_k}}, & \text{if ${\beta}_{{c_i},{r_k}} < {\gamma}_{{s_j},{r_k}}$} \\
			0. & \text{otherwise}
		\end{array}
		\right.
	\end{equation}
	The procedure of the DP algorithm is shown in Algorithm \ref{alg:dp}.
	
	FFD is a greedy algorithm in which the containers are sorted in decreasing order according to their individual weight that is determined by each container's demands for multi-type resources, and then containers are placed sequentially in the first server that has sufficient capacity to accommodate them. %由于每个容器所需要的资源是多维的，在多维情况下对尺寸的定义决定了容器的放置顺序。应用程序对资源的需求是不同的，我们将应用程序需求最多的资源作为排序的依据。容器按照RK的大小进行非增排列，遍历服务器，容器放置在第一个可以满足其需求的服务器上。
	As the resources required by each container are multidimensional (i.e., multi-type), the definition of dimensions and the associated weight assigned to the vector of resources required by an individual container determine the order in which the containers are placed. Containers have different resource requirements and there are multiple alternative methods for selecting an appropriate weight for each container. For example, one can calculate the weighted sum of multi-type resources required by the container as the basis for the decreasing ordering. In particular, when the weights are equal, containers are decreasingly ordered by evaluating $\sum_{k=1}^{R} {{\beta}_{{c_j},{r_k}}}$, $c_j\in\mathcal{C}$. 
	%$\mathop {\max}_j\{\sum_{k=1}^{R} {{\beta}_{{c_j},{r_k}}|c_j\in\mathcal{C}}\}$
	Additionally, if the demands of a particular type of resource always dominate the demands of the other types of resources, one can only consider the dominant type of resource to determine the weights of containers and the decreasing order. Then, upon traversing the servers, the containers are sequentially  placed on the first server that can satisfy their requirements of resources. The details of FFD are presented in Algorithm \ref{alg:ffd}.
	
	To sum up, the process of the whole task containerization and container placement scheme is as follows: An application's tasks are first grouped by the P-NCPI or the P-RI algorithm, and then each group of tasks, denoted by ${\mathcal A}_i$, is encapsulated into a container $c_i$, which is then placed on the selected server by using the DP or the FFD algorithm. In this manner, the application is eventually executed in a given number of containers and on the selected servers. As a beneficial result, the proposed scheme exhibits minimized inter-container communication cost (some containers are possibly placed on different servers), balanced resource requirements among containers, and balanced resource utilization efficiency among the selected servers.
	         
	\begin{algorithm}
		\small
		\caption{The proposed task containerization algorithm}
		\label{alg:tca}
		\renewcommand{\algorithmicrequire}{\textbf{Input:}}
		\renewcommand{\algorithmicensure}{\textbf{Output:}}
		\begin{algorithmic}[1]
			\REQUIRE $\mathcal{G}=(\mathcal{T}, \mathcal{E})$, $\mathcal{R}$, $C$, $\{\mathbf{v}_i\}_{i=1}^{|\mathcal{T}|}$    
			\ENSURE  $\mathcal{P}$, $\mathbf{D}$
			{   \STATE Initialization: $\mathcal{P} = ({{{\mathcal A}_1}=\emptyset,{{\mathcal A}_2}=\emptyset, \cdots ,{{\mathcal A}_C}=\emptyset})$
				\IF{RI is the initial partitioning algorithm}
				\STATE select $C$ tasks in $\mathcal{T}$ randomly 					
				\STATE allocate each of the $C$ tasks exclusively to all $\mathcal{A}_i$, $i=1, 2, \cdots, C$ 
				\ENDIF}
			\IF {{NCPI  is the initial partitioning algorithm}}
			\STATE $\bf z \gets$ sort vertices in $\mathcal{G}$ topologically
			\IF{$\mathcal{G}$ has a loop}
			\STATE raise Error
			\ENDIF
			\STATE Generate a $|{\bf z}|$-dimensional null vector $\mathbf{h}$
			\FOR{$i = 1$ to $|\mathcal{T}|$}
			%\STATE $ptr\gets topo[i]$
			\STATE ${\mathcal N}_{z_i}\gets$ the indices of the adjacency vertices of $z_i$
			\FOR{$j=1$ to $|\mathcal{N}_{z_i}|$}
			\IF{$h_{\mathcal{N}_{z_i}[j]} < h_{\mathcal{N}_{z_i}[j]} + \omega(e_{\mathcal{N}_{z_i}[j],z_i})$}
			\STATE$h_{\mathcal{N}_{z_i}[j]} \gets h_{\mathcal{N}_{z_i}[j]} + \omega(e_{\mathcal{N}_{z_i}[j],z_i})$
			\ENDIF
			\ENDFOR
			\ENDFOR
			\STATE Initialize a vector $\mathbf{u}$ with all elements being ${\mathop {\max}\{h_i|h_i\in\mathbf{h}\}}$
			\STATE Reverse the elements order of $\bf z$ 	                \STATE Repeat Lines 12-19 to obtain $\mathbf{u}$
			\FOR{$i = 1$ to $|\mathcal{T}|$}
			%\STATE $ptr\gets topo[i]$
			\STATE ${\mathcal N}_{i}\gets$ the indices of the adjacency vertices of vertex $i$
			\FOR{$j =1$ to $|\mathcal{N}_i|$ }
			\STATE$l \gets h_i$
			\STATE$v \gets u_i-\omega(e_{\mathcal{N}_i[j],i})$
			\IF{$l == v$}
			\STATE include $e_{\mathcal{N}_i[j],i}$ into the critical path
			\ENDIF
			\ENDFOR
			\ENDFOR
			%\STATE Find the critical path
			\STATE Select $C$ tasks on the non-critical path
			\STATE Allocate each of the $C$ tasks exclusively to all $\mathcal{A}_i$, $i=1, 2, \cdots, C$
			\ENDIF
			\STATE Put the remaining $|\mathcal{T}|-C$ tasks into the set $\mathcal{Y}$
			\FOR{$j = 1$ to $|\mathcal{T}|-C$}
			\STATE score $\gets \ 0 $
			\FOR{$i=1$ to $C$}
			\STATE calculate $\Delta h({\mathcal Y}[j],{{\mathcal A}_i})$ by Equation (\ref{inter})
			\IF{$\Delta h({\mathcal Y}[j],{{\mathcal A}_i}) > $ score}
			\STATE score $\leftarrow \Delta h({\mathcal Y}[j],{{\mathcal A}_i})$
			\STATE $k \leftarrow i$
			\ENDIF
			\ENDFOR
			\STATE Put task ${\mathcal Y}[j]$  into ${\mathcal A}_k$
			\ENDFOR
			\STATE Generate $\mathbf{D}$ according to $\mathcal{P}$
		\end{algorithmic}
	\end{algorithm}
	
	\begin{algorithm}
		\small
		\caption{The DP algorithm}
		\label{alg:dp}
		\renewcommand{\algorithmicrequire}{\textbf{Input:}}
		\renewcommand{\algorithmicensure}{\textbf{Output:}}
		\begin{algorithmic}[1]
			\REQUIRE $|\mathcal{T}|$, $\mathcal{C}$, $\mathcal{S}$, $\mathcal{R}$, $\{\mathbf{v}_i\}_{i=1}^{|\mathcal{T}|}$, $\mathbf{D}$
			\ENSURE  $\mathbf{M}$
			\STATE Initialization: Generate a null matrix $\mathbf{M}={\bf 0}_{|\mathcal{C}| \times |\mathcal{S}|}$
			\FOR{$i=1$ to $|\mathcal{C}|$}
			\FOR{$j=1$ to $|\mathcal{S}|$}
			\IF{server ${s_j}$ can provide all the resources required by container ${c_i}$}
			\STATE $\textsf{DP}_{ij} \gets$ calculate the dot product of multi-type resources required by container ${c_i}$ and multi-type resources that server ${s_j}$ can provide
			\ELSE 
			\STATE $\textsf{DP}_{ij} \gets 0$
			\ENDIF
			\ENDFOR
			\STATE Select the server ${s_j}$ with the largest
			$\textsf{DP}_{ij}$ to accommodate the container ${c_i}$
			\STATE $m_{ij} \gets 1$
			\ENDFOR
		\end{algorithmic}
	\end{algorithm}

	\begin{algorithm}
		\small
		\caption{The FFD algorithm}
		\label{alg:ffd}
		\renewcommand{\algorithmicrequire}{\textbf{Input:}}
		\renewcommand{\algorithmicensure}{\textbf{Output:}}
		\begin{algorithmic}[1]
			\REQUIRE $|\mathcal{T}|$, $\mathcal{C}$, $\mathcal{S}$, $\mathcal{R}$, $\{\mathbf{v}_i \}_{i=1}^{|\mathcal{T}|}$, $\mathbf{D}$ 
			\ENSURE  $\mathbf{M}$
			\STATE Initialization: Generate a null matrix $\mathbf{M}={\bf 0}_{|\mathcal{C}| \times |\mathcal{S}|}$
			\STATE Sorts the containers in decreasing order by evaluating $\sum_{k=1}^{R} {{\beta}_{{c_j},{r_k}}}$ of each container $c_j$
			\FOR{$i = 1$ to $|\mathcal{C}|$}
			\FOR{$j=1$ to $|\mathcal{S}|$}
			\IF{server ${s_j}$ can provide all the resources required by container ${c_i}$}
			\STATE select server ${s_j}$ to place container ${c_i}$
			\STATE $m_{ij} \gets 1$
			\ENDIF
			\IF{$m_{ij} == 1$}
			\STATE break
			\ENDIF
			\ENDFOR
			\ENDFOR 
		\end{algorithmic}
	\end{algorithm}
	
	\subsection{Analysis of Algorithm Complexity}
	The task containerization method groups the tasks of the given application into multiple vertex sets, each of which is encapsulated into a single container. Then the generated containers are placed on the selected appropriate servers by using the container placement method. 
	
	More specifically, two initial partitioning algorithms, i.e., P-NCPI and P-RI, are individually employed in the task containerization method. P-RI initializes a given number of vertex sets, which are supposed to represent the task partitioning, by using the same number of randomly selected tasks (see Lines 2-5 of Algorithm \ref{alg:tca}). The time complexity of P-RI is $O \left(T^{2} +2T \right) = O \left(C \right) +  O \left(C \right) + O \left(T-C \right)+ O \left(T(T-C) \right) +O \left(T-C \right) + O \left(TC \right)$, and the number of floating point operations is given by $1+C+C+T-C+T-C+2T(T-C)+T-C+TC = 2T^2-TC+3T-C+1$. P-NCPI employs the non-critical path based initialization to conduct the graph partitioning (see Lines 6-35 of Algorithm \ref{alg:tca}). Its time complexity is $O \left(4T^{2} +2T  \right) =  O \left(3T^2 \right) +  O \left(2C \right) + O \left(T-C \right)+ O \left(T(T-C) \right) +O \left(T-C \right) + O \left(TC \right) $, and the number of floating point operations is given by $1+T^2+1+1+T+T^2+1+T+T^2+C+C+T-C+T-C+2T(T-C)+T-C+TC = 5T^2-TC+5T-C+4$. 
	
	As far as the container placement method is concerned, two bin packing algorithms are used, namely DP (see Algorithm \ref{alg:dp}) and FFD (see Algorithm \ref{alg:ffd}). The time complexity of DP is $O\left ( CS \right )$ and the number of floating point operations is $1+CS+2C$. The time complexity of FFD is $O\left ( CR+CS + C\right )$, and the number of floating point operations is $1+CR+2CS+C$. 
	
	\section{Evaluation}\label{sec5}
	In this section, we present simulation results to demonstrate the effectiveness of our proposed task containerization and container placement methods. The experiments were carried out on a MacBook Pro with a 4-core CPU, 8GB of RAM, and the macOS Monterey operating system. All the algorithms presented in this paper were implemented by Python 3.9.7. For the P-RI algorithm and the $K$-means algorithm, we took the average of 10 sets of experimental results to obtain statistically reliable results. Moreover, all the results were calculated relying on the experiments of 10 application programs, each of which is represented by a workflow composed of a different number of tasks: 5, 9, 17, 24, 30, 47, 57, 63, 91  and 93.
	
	We considered the CPU and memory resources of a computing network comprising 10 MEC servers. The server configurations are listed in Table \ref{server}, where the volume of each type of computing resources is normalized. We set the parameters $\mu_{C}$ as 0.5, $\mu_{B}$  as 0.5, and $\theta$ as 1.5.	The used data regarding the application tasks were obtained from the Alibaba Cloud platform \cite{ALI}.  Each container hosts at least one process, and each task is set up as a separate process. In general, resources allocated to a container are more than the container needs \cite{SPR}. We ignored the discrepancy between the volume of resources allocated to the container and the volume of resources demanded by the container for the convenience of statistics and visualization. 
	
	%We set each task as a separate process, and the container contains one or more processes. Generally speaking, containers that host application tasks need to set a certain usage quota. These resource quotas usually require more resources than the application needs. For the convenience of statistics and representation, we ignore the resource difference between the resources allocated to the container and the container demand. 
	
	%We simulate a small edge computing platform with 10 servers considering CPU and memory resources. Table \ref{server} lists the server configurations for each resource after normalization.

	\begin{table}[tb]
    \centering \caption{{Configuration of MEC servers}}\label{server}
   \setlength{\tabcolsep}{7mm}{
	\begin{tabular}{ccc}
			\toprule
			\textbf{Server}& \textbf{CPU (\%)}& {\textbf{Memory (\%)}}  \\ \midrule
			0& 7& 7  \\ 
			1& 9& 8\\ 
			2& 10& 8\\ 
			3& 12& 11\\ 
			4&6 & 11\\ 
			5& 12& 14\\ 
			6& 14& 8\\
			7& 10& 11\\ 
			8& 9& 14\\ 
			9& 11& 8 \\
           \bottomrule
           \end{tabular}}
           \end{table}

	%We compare the proposed task partition approach with the clustering algorithm K-means. 
	%We compare the proposed method with state-of-the-art container distribution strategy used by container platform provider docker swarm, as well as k-means.
	%K-means: Randomly select $l$ objects as the initial cluster center, then calculate the distance between each object and each seed cluster center, and assign each object to the nearest cluster center to it.
	
	%For the joint task containerization and container placement solution, we compare it with the state-of-the-art container allocation strategy Spread used by container platform provider Docker swarm.
	
	%这两个对比算法思路如下：
	%Spread: Preference is given to the node with the least occupied resources so that the resources of all nodes in the cluster are used equally.
	
	\begin{figure}[t]
		
		\includegraphics[width = 3.5in]{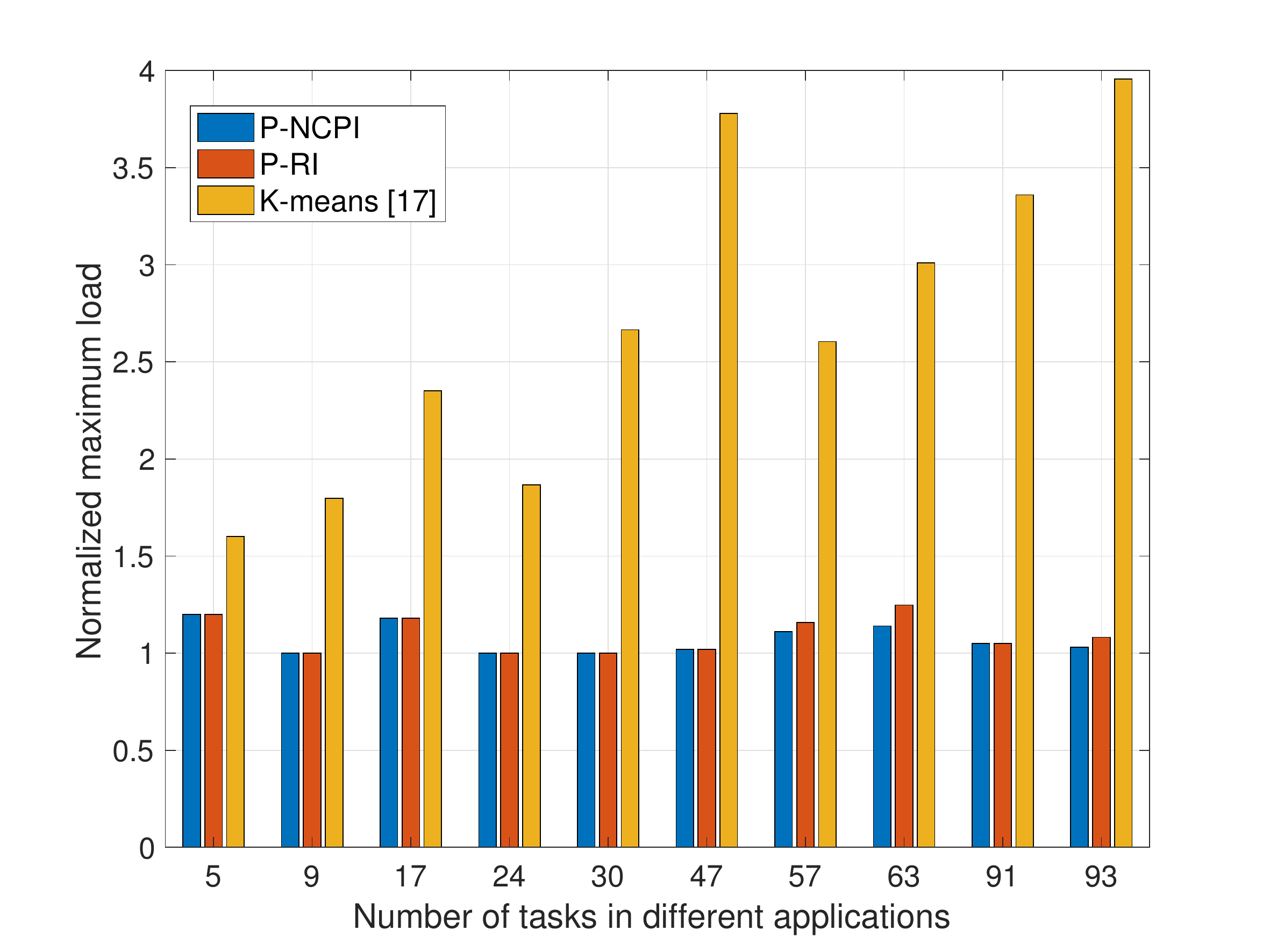}
		\caption{Comparison of the normalized maximum load performance (defined by Equation \eqref{normalized_maximum_load}) of various task partitioning algorithms, including the proposed P-NCPI and P-RI algorithms, and the classic $K$-means algorithm.}
		\label{fig:load}
	\end{figure}		
	The balance degree of utilizing computing resources is a critical performance indicator for MEC. To avoid the problem of long request delay caused by insufficient resources of the servers, the utilization of multi-type computing resources on the individual servers has to be balanced. In Figure~\ref{fig:load}, we compared the normalized maximum load $\lambda$ (defined by Equation  \eqref{normalized_maximum_load}) of the proposed task partitioning algorithms and of the benchmarking $K$-means clustering method. Specifically, we observe in Figure~\ref{fig:load} that the normalized maximum load $\lambda$ of the $K$-means method \cite{JAH} varies greatly between 1.6 and 4. Unlike those of the $K$-means method, the values of $\lambda$ for P-NCPI and P-RI remain between 1.25 and 1, indicating that the resource requirements by each group of tasks are more balanced after conducting task partitioning with our proposed algorithms (e.g., decreasing the normalized maximum load by up to 60.24\% on average when using P-NCPI). 
	
	\begin{figure}[t]
		\includegraphics[width = 3.5in]{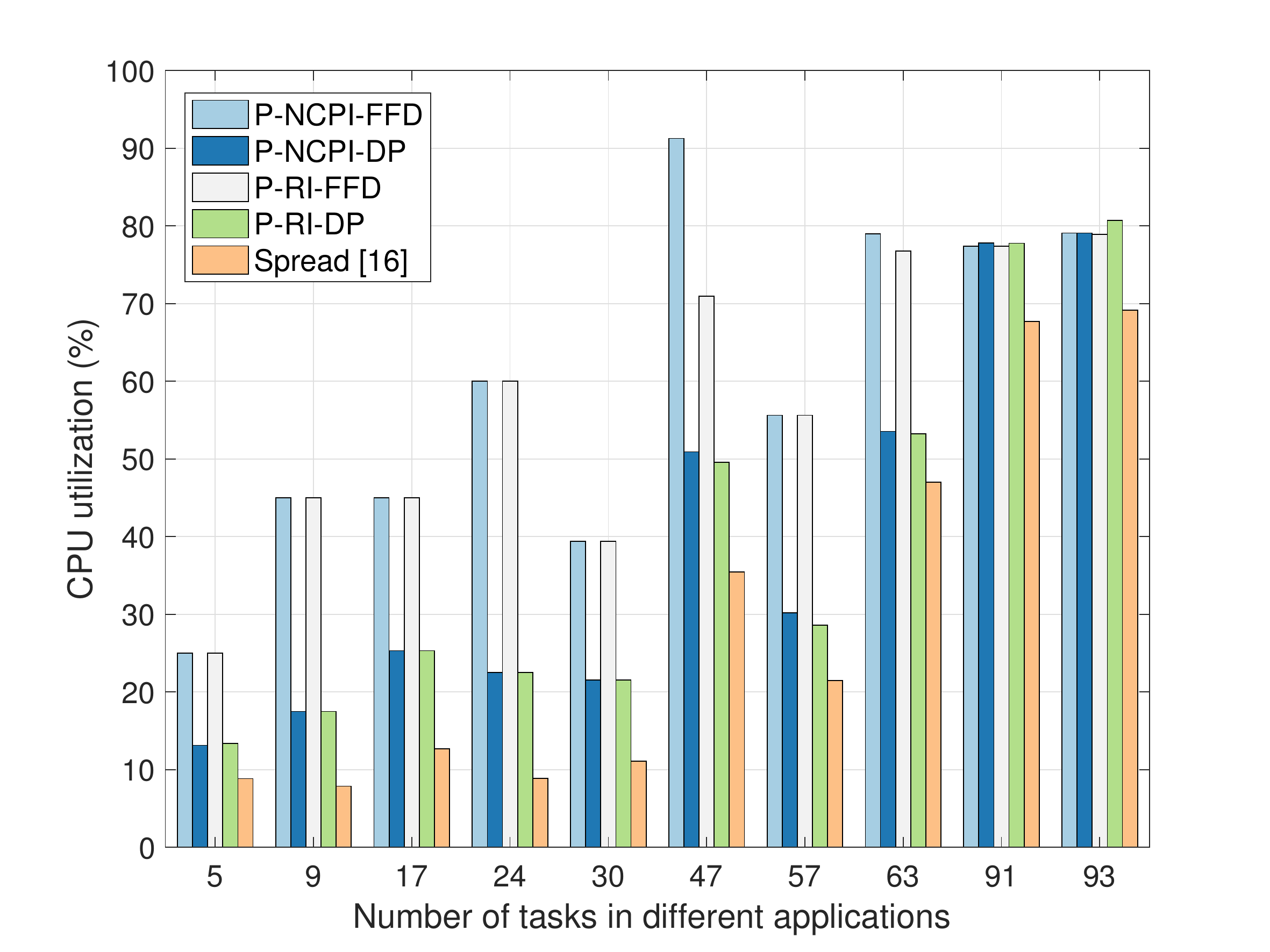}
		\caption{Comparison of the average CPU utilization efficiency of the MEC-based computing network (defined by Equation \eqref{CPU_utilization}), when using the proposed four task-containerization-and-container-placement algorithms and the Spread algorithm.}
		\label{fig:cpu}
	\end{figure}
	%\vspace{-10pt}
	\begin{figure}[t]
		\includegraphics[width = 3.5in]{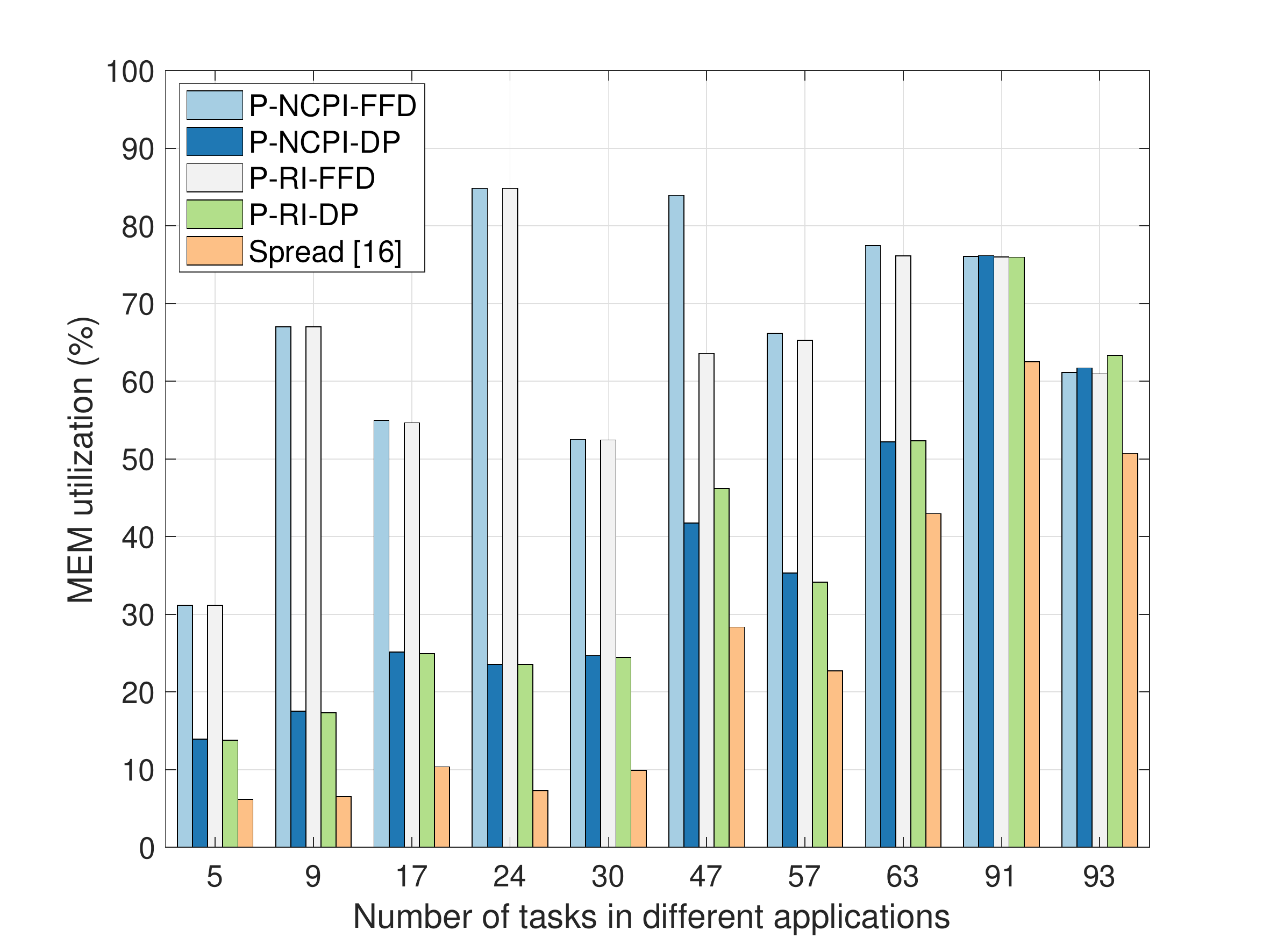}
		\caption{Comparison of the average MEM utilization efficiency of the MEC-based computing network (defined by Equation \eqref{MEM_utilization}), when using the proposed four task-containerization-and-container-placement algorithms and the Spread algorithm.}
		\label{fig:mem}
	\end{figure}
	
	In Figures \ref{fig:cpu} and \ref{fig:mem}, we compared the proposed task-containerization-and-container-placement methods with the state-of-the-art strategy ``Spread'' \cite{SPR} employed by the container orchestration tool Docker swarm, in terms of the average CPU and the memory (i.e., MEM) utilization efficiencies in the MEC-based computing network. Specifically, the average CPU utilization efficiency of the MEC-based computing network is defined as 
	\begin{equation}\lambda_{\textrm{CPU}} = \frac{\sum_n^{N_\textrm{S}} \eta_n}{N_\textrm{S}},\label{CPU_utilization}
	\end{equation}where $N_\textrm{S}$ represents the number of servers occupied by the application, $\eta_n$ denotes the CPU utilization efficiency of server $n$. Similarly, the average MEM utilization efficiency of the MEC-based computing network is defined as 
	\begin{equation}
		\lambda_{\textrm{MEM}} = \frac{\sum_n^{N_\textrm{S}} \zeta_n}{N_\textrm{S}},\label{MEM_utilization}
	\end{equation}where $\zeta_n$ represents the MEM utilization efficiency of server $n$. 
	
	We see from Figures \ref{fig:cpu} and \ref{fig:mem} that the proposed P-NCPI-FFD algorithm improves the average CPU utilization efficiency by 30.66\% and the average MEM utilization efficiency by 40.77\%, compared with the Spread algorithm. In general, the proposed P-NCPI-FFD algorithm achieves the highest CPU and MEM utilization efficiency among all algorithms considered. The FFD algorithm attains higher resources utilization efficiency than the DP algorithm. This phenomenon can be explained as follows. The DP algorithm focuses on finding out the largest sum product of a container's multi-type resource requirements and the volume of multi-type resources that a server can provide, whereas FFD puts as many containers having higher resource requirements as possible on the minimum possible number of servers. Therefore, when using FFD, $\eta_n$ and $\zeta_n$ in Equations \eqref{CPU_utilization} and \eqref{MEM_utilization} are larger, while $N_\textrm{S}$ is smaller, than those when employing DP. In addition, it is observed that the impact of the P-NCPI and P-RI algorithms on the results of CPU and MEM utilization is trivial.	
	
\begin{figure}[t]
		\includegraphics[width = 3.5in]{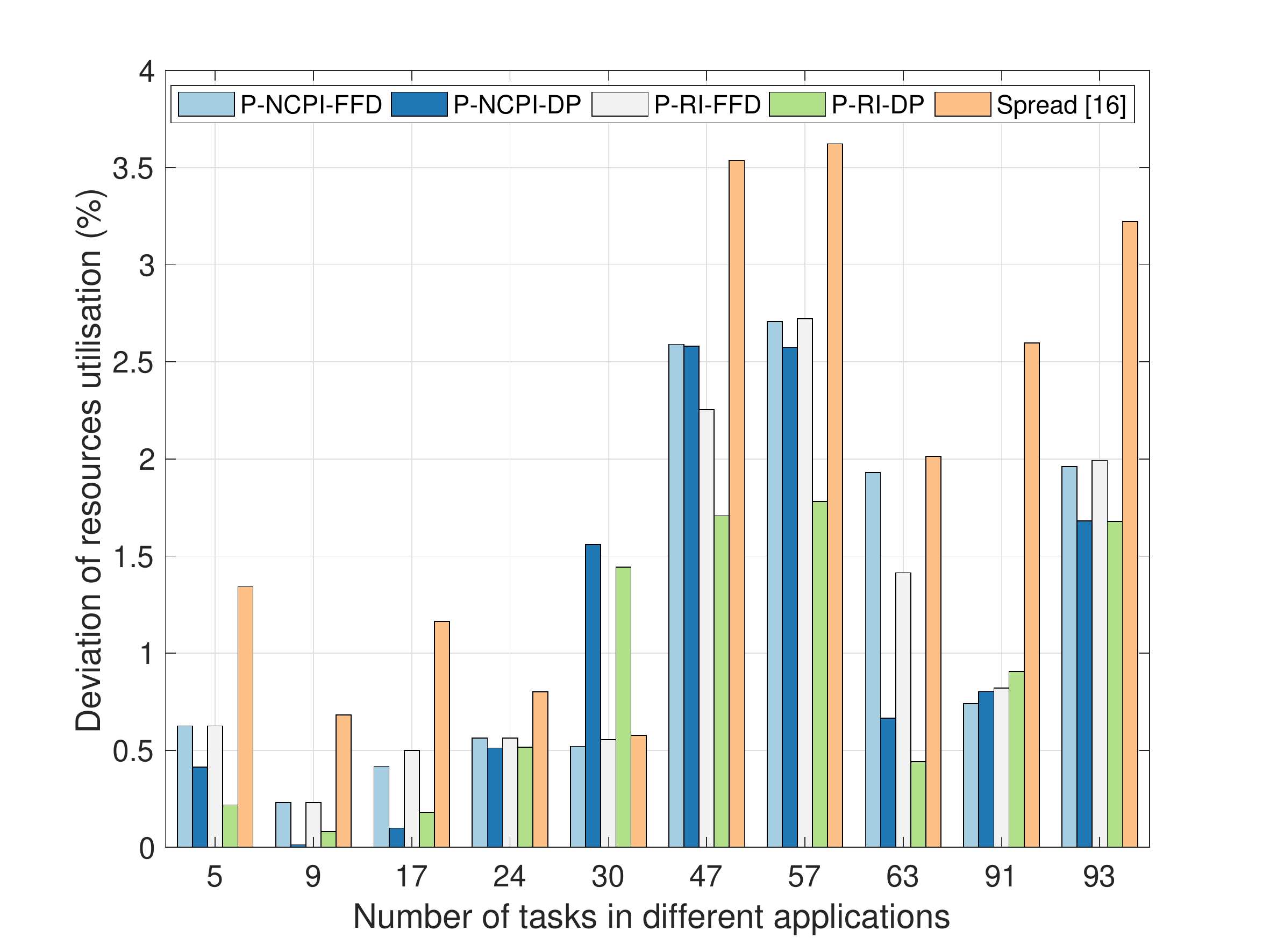}
		\caption{Comparison of the balance degree of multi-type computing resource utilization (defined by Equation \eqref{total_deviation}), when using the proposed four task-containerization-and-container-placement algorithms and the Spread algorithm.}
		\label{fig:var}
	\end{figure}

	For a given application, the metric defined by Equation \eqref{total_deviation} can reflect the extent to which different resources are used in a balanced manner. According to Figure \ref{fig:var}, in most cases evaluated, our proposed algorithms outperform the Spread algorithm in terms of the balance degree of utilizing multi-type computing resources. Although the DP algorithm is generally inferior to the FFD algorithm in terms of the CPU and MEM utilization efficiencies (as shown by Figure \ref{fig:cpu} and Figure \ref{fig:mem}), it performs better than FFD in terms of the balance degree of utilizing multi-type computing resources across the MEC-based computing network. The P-NCPI and P-RI algorithms, which are invoked for the initial task partitioning, do not make much difference in this respect.
	
\begin{figure}[t]
		\includegraphics[width = 3.5in]{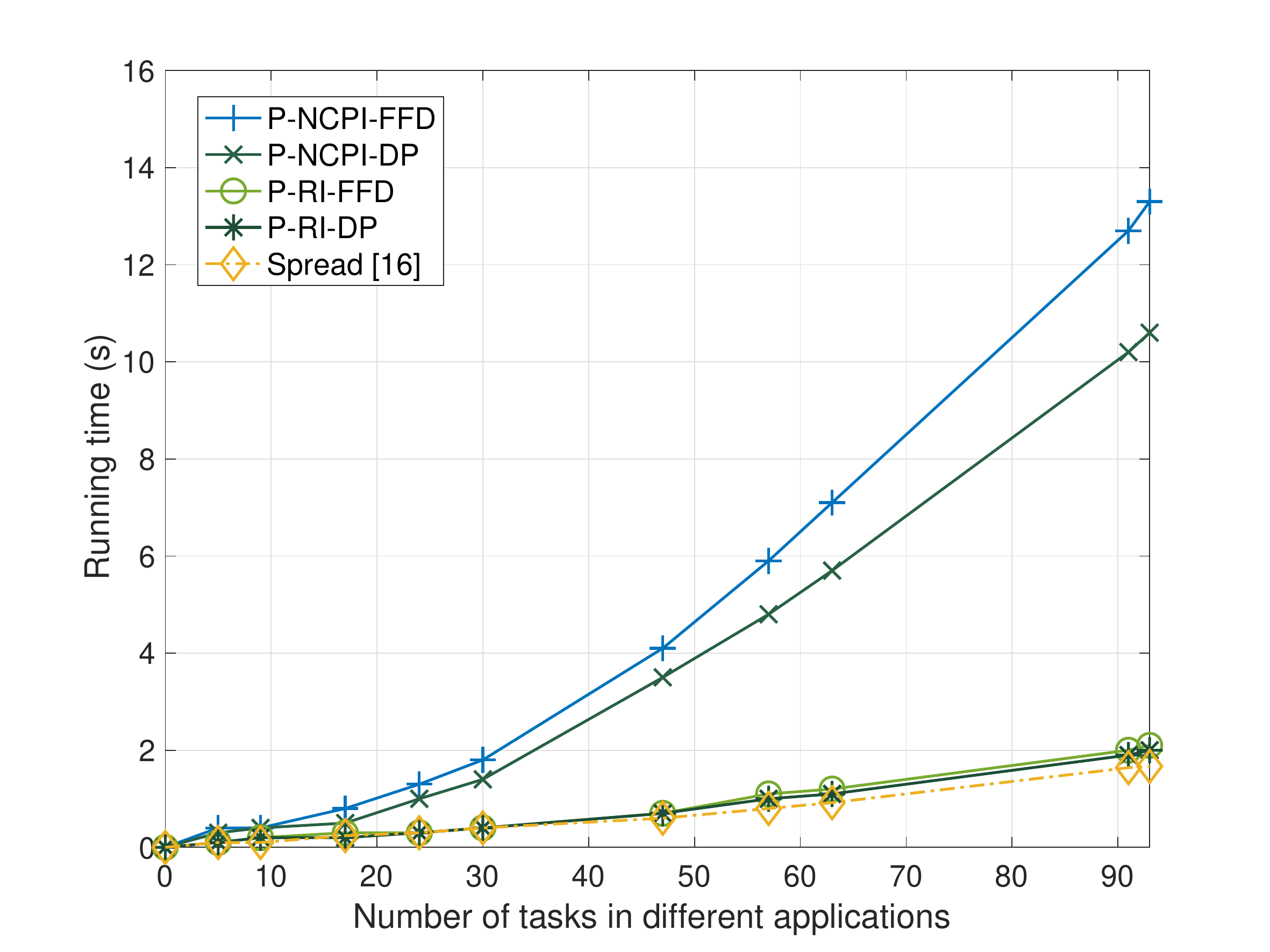}
		\caption{Comparison of the running time of different task-containerization-and-container-placement algorithms.}
		\label{fig:runtime}
	\end{figure}

	The running time results of all the task-containerization-and-container-placement algorithms considered are provided in Figure \ref{fig:runtime}. It is observed that the running time of the P-RI algorithm is shorter than that of the P-NCPI algorithm. Additionally, when the number of tasks exceeds a particular value, the DP algorithm has a shorter running time than the FFD algorithm. Among the four algorithms we proposed, the P-RI-DP algorithm has the shortest running time, and in this regard, there is very little difference between the P-RI-DP algorithm and the Spread algorithm. Notably, the Spread algorithm always performs best in terms of the running time. 
	%服务器之间通信的时间占应用程序总运行时间的百分比用于计算边缘计算平台上的通信成本
	
\begin{figure}[t]
		\includegraphics[width = 3.5in]{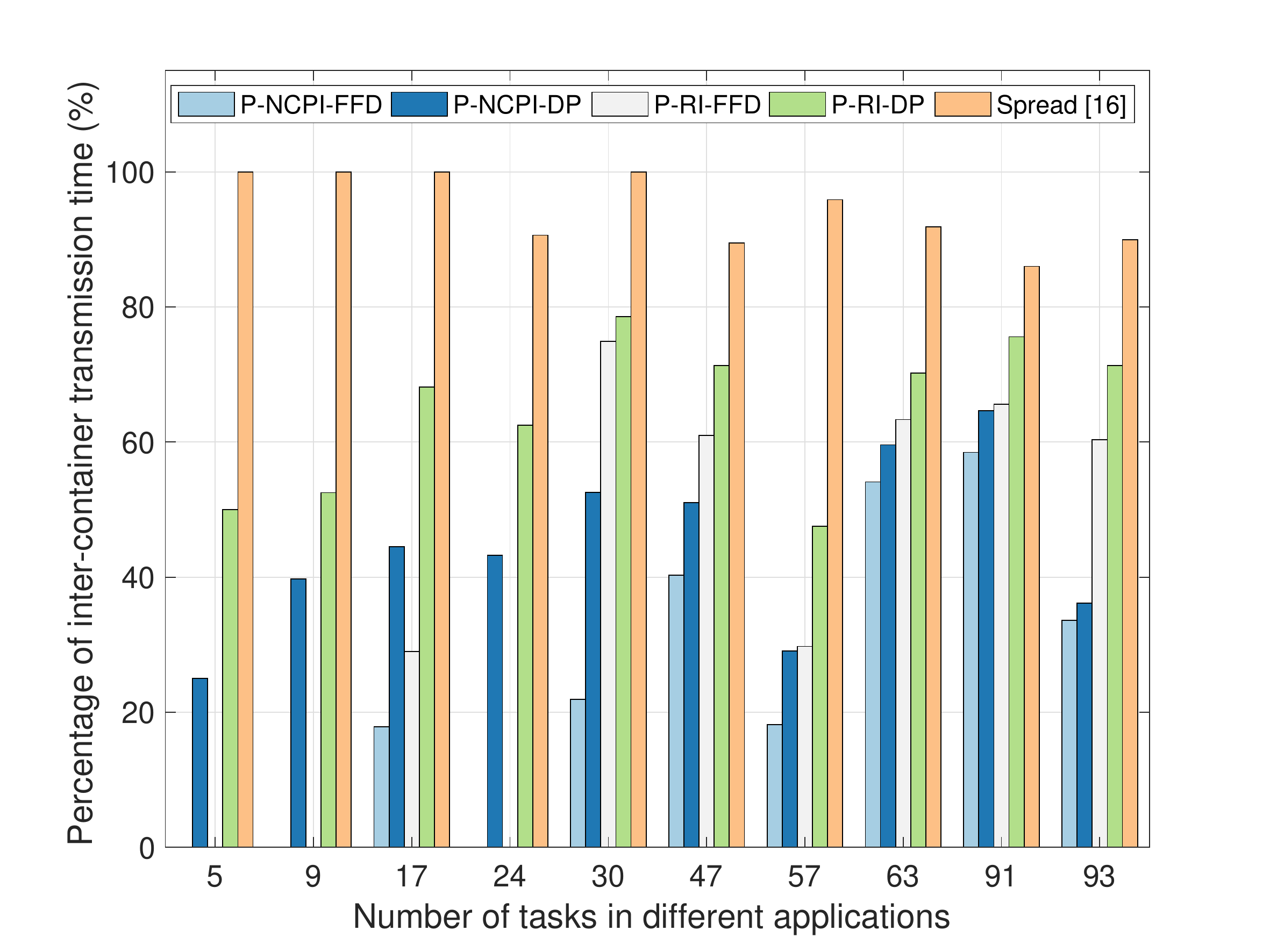}
		\caption{The ratio of the inter-container communication overhead to the inter-task communication overhead when using different task-containerization-and-container-placement algorithms.}
		\label{fig:interData}
	\end{figure}

	Finally, the results of the communication overhead for various task-containerization-and-container-placement algorithms are shown in Figure \ref{fig:interData}. We can see that when compared with the benchmarking Spread algorithm, the P-NCPI-FFD algorithm saves 74.10\% of the communication overhead, while the P-RI-FFD algorithm saves 59.32\%. The communication overhead of FFD is lower than that of DP, because it does not have to consider the balance between the resource demand of containers and the resource supply of servers.
	
	To sum up, our proposed task-containerization-and-container-placement algorithms outperform the existing algorithms in general. In particular, all of our proposed algorithms exhibit a more balanced resource requirements by containers, the P-NCPI-FFD algorithm achieves the highest CPU and MEM utilization efficiencies, the P-NCPI-DP and P-RI-DP algorithms achieve a higher degree of balance in utilizing multi-type computing resources, while the P-NCPI-FFD and P-RI-FFD algorithms perform better in terms of communication overhead. Additionally, different initial task partitioning algorithms, such as P-NCPI and P-RI, do not make much difference in terms of the above performance metrics.  
	
	\section{Conclusions}\label{sec6}
	In this paper we have proposed four task-containerization-and-container-placement algorithms to jointly reduce the inter-container communication overhead and balance the multi-type computing resource utilization in the MEC-based computing network.
	We establish a workflow model to reflect the interactions between interdependent tasks. As a beneficial result, the inter-task communication overhead, the inter-container communication overhead, as well as the computing resources required by each group of tasks encapsulated in a container, are conveniently characterized. Furthermore, two task containerization algorithms have been designed to reduce the inter-container communication overhead while balancing the multi-type computing resource requirements of containers. Then we proposed a pair of container placement algorithms for optimizing the utilization of multi-type computing resources across MEC servers. Extensive simulation results demonstrated that our proposed methods are capable of reducing the inter-container communication overhead by up to 74.10\%, reducing the normalized maximum load by up to 60.24\%, improving the CPU utilization efficiency by up to 30.66\%, and improving the memory utilization efficiency by up to 40.77\% in the MEC-based computing network considered.

\end{document}